\documentclass[prl,twocolumn,superscriptaddress,amsmath,amssymb,aps,floatfix]{revtex4-2}
\usepackage{soul}
\usepackage[colorlinks=true,citecolor=blue,filecolor=blue,linkcolor=blue,urlcolor=blue,pdftex]{hyperref}
\usepackage{graphicx}
\usepackage{dcolumn}
\usepackage{bm}
\usepackage[nameinlink,capitalize]{cleveref}
\newcommand{\bologna}{\affiliation{Department of Physics and Astronomy, University of Bologna and INFN-Bologna, 40126 Bologna, Italy}}
\newcommand{\chicago}{\affiliation{Department of Physics \& Kavli Institute for Cosmological Physics, University of Chicago, Chicago, IL 60637, USA}}
\newcommand{\coimbra}{\affiliation{LIBPhys, Department of Physics, University of Coimbra, 3004-516 Coimbra, Portugal}}
\newcommand{\columbia}{\affiliation{Physics Department, Columbia University, New York, NY 10027, USA}}
\newcommand{\lngs}{\affiliation{INFN-Laboratori Nazionali del Gran Sasso and Gran Sasso Science Institute, 67100 L'Aquila, Italy}}
\newcommand{\mainz}{\affiliation{Institut f\"ur Physik \& Exzellenzcluster PRISMA$^{+}$, Johannes Gutenberg-Universit\"at Mainz, 55099 Mainz, Germany}}
\newcommand{\heidelberg}{\affiliation{Max-Planck-Institut f\"ur Kernphysik, 69117 Heidelberg, Germany}}
\newcommand{\munster}{\affiliation{Institut f\"ur Kernphysik, Westf\"alische Wilhelms-Universit\"at M\"unster, 48149 M\"unster, Germany}}
\newcommand{\nikhef}{\affiliation{Nikhef and the University of Amsterdam, Science Park, 1098XG Amsterdam, Netherlands}}
\newcommand{\nyuad}{\affiliation{New York University Abu Dhabi - Center for Astro, Particle and Planetary Physics, Abu Dhabi, United Arab Emirates}}
\newcommand{\purdue}{\affiliation{Department of Physics and Astronomy, Purdue University, West Lafayette, IN 47907, USA}}
\newcommand{\rice}{\affiliation{Department of Physics and Astronomy, Rice University, Houston, TX 77005, USA}}
\newcommand{\stockholm}{\affiliation{Oskar Klein Centre, Department of Physics, Stockholm University, AlbaNova, Stockholm SE-10691, Sweden}}
\newcommand{\subatech}{\affiliation{SUBATECH, IMT Atlantique, CNRS/IN2P3, Universit\'e de Nantes, Nantes 44307, France}}
\newcommand{\torino}{\affiliation{INAF-Astrophysical Observatory of Torino, Department of Physics, University  of  Torino and  INFN-Torino,  10125  Torino,  Italy}}
\newcommand{\ucsd}{\affiliation{Department of Physics, University of California San Diego, La Jolla, CA 92093, USA}}
\newcommand{\wis}{\affiliation{Department of Particle Physics and Astrophysics, Weizmann Institute of Science, Rehovot 7610001, Israel}}
\newcommand{\zurich}{\affiliation{Physik-Institut, University of Z\"urich, 8057  Z\"urich, Switzerland}}
\newcommand{\paris}{\affiliation{LPNHE, Sorbonne Universit\'{e}, CNRS/IN2P3, 75005 Paris, France}}
\newcommand{\freiburg}{\affiliation{Physikalisches Institut, Universit\"at Freiburg, 79104 Freiburg, Germany}}
\newcommand{\napels}{\affiliation{Department of Physics ``Ettore Pancini'', University of Napoli and INFN-Napoli, 80126 Napoli, Italy}}
\newcommand{\nagoya}{\affiliation{Kobayashi-Maskawa Institute for the Origin of Particles and the Universe, and Institute for Space-Earth Environmental Research, Nagoya University, Furo-cho, Chikusa-ku, Nagoya, Aichi 464-8602, Japan}}
\newcommand{\laquila}{\affiliation{Department of Physics and Chemistry, University of L'Aquila, 67100 L'Aquila, Italy}}
\newcommand{\tokyo}{\affiliation{Kamioka Observatory, Institute for Cosmic Ray Research, and Kavli Institute for the Physics and Mathematics of the Universe (WPI), University of Tokyo, Higashi-Mozumi, Kamioka, Hida, Gifu 506-1205, Japan}}
\newcommand{\kobe}{\affiliation{Department of Physics, Kobe University, Kobe, Hyogo 657-8501, Japan}}

\newcommand{\kit}{\affiliation{Institute for Astroparticle Physics, Karlsruhe Institute of Technology, 76021 Karlsruhe, Germany}}
\newcommand{\tsinghua}{\affiliation{Department of Physics \& Center for High Energy Physics, Tsinghua University, Beijing 100084, China}}
\newcommand{\ferrara}{\affiliation{INFN - Ferrara and Dip. di Fisica e Scienze della Terra, Universit\`a di Ferrara, 44122 Ferrara, Italy}}

\begin{document}
\preprint{APS/123-QED}

\author{E.~Aprile}\columbia
\author{K.~Abe}\tokyo
\author{S.~Ahmed Maouloud}\paris
\author{L.~Althueser}\munster
\author{B.~Andrieu}\paris
\author{E.~Angelino}\torino
\author{J.~R.~Angevaare}\nikhef
\author{V.~C.~Antochi}\stockholm
\author{D.~Ant\'on Martin}\chicago
\author{F.~Arneodo}\nyuad
% ...

\collaboration{XENON Collaboration}
\email[]{xenon@lngs.infn.it}
\noaffiliation
\title{Searching for Heavy Dark Matter near the Planck Mass with XENON1T}

\begin{abstract}
Multiple viable theoretical models predict heavy dark matter particles with a mass close to the Planck mass, a range relatively unexplored by current experimental measurements.  We use 219.4~days of data collected with the XENON1T experiment to conduct a blind search for signals from Multiply-Interacting Massive Particles (MIMPs). Their unique track signature allows a targeted analysis with only 0.05~expected background events from muons. Following unblinding, we observe no signal candidate events. This work places strong constraints on spin-independent interactions of dark matter particles with a mass between 1$\times$10$^{12}$\,GeV/c$^2$ and 2$\times$10$^{17}$\,GeV/c$^2$. In addition, we present the first exclusion limits on spin-dependent MIMP-neutron and MIMP-proton cross-sections for dark matter particles with masses close to the Planck scale.
\end{abstract}

\maketitle

Despite numerous pieces of evidence for dark matter (DM) and decades of dedicated searches, the nature of DM remains a mystery~\cite{Billard:2021uyg}. 
A wide class of production mechanisms predicts DM candidates near the Planck mass ($\simeq$10$^{19}$\,GeV/c$^2$), such as non-standard thermal freeze-out~\cite{Chu:2011be,Berlin:2016gtr}, thermal freeze-in~\cite{Chung:1998rq,Giudice:2000ex}, first-order phase transitions~\cite{Azatov:2021ifm,Baldes:2021aph}, decays of heavy fields~\cite{Takahashi:2007tz,Acharya:2009zt}, gravitational particle production~\cite{Chung:1998zb,Ema:2019yrd}, and primordial black holes~\cite{MacGibbon:1987my,Maldacena:2020skw}. 
DM candidates in this mass range are less constrained by experiments and too heavy to be produced at colliders. 
 
There has been a growing interest to search for such heavy DM candidates~\cite{Carney:2022gse}, where new detection constraints have been explored considering spin-independent~(SI) scattering. Assuming coherent interaction with a nucleus of mass number $A$, there is an $A^4$\textendash enhanced sensitivity in SI scattering~\cite{Goodman:1984dc,Digman:2019wdm}. Xenon, with its large mass number, particularly benefits from this enhancement. However, the spin-dependent (SD) channel lacks experimental constraints with the exception of an early study~\cite{Bernabei:1999ui}. Xenon experiments can set leading SD constraints using the naturally abundant isotopes with nonzero nuclear spin~\cite{Jungman:1995df,XENON:2019rxp}. 

Under the assumption of the Standard Halo Model~\cite{Baxter:2021pqo}, the total DM flux decreases with particle mass due to the fixed local DM mass density~\cite{Read:2014qva}, which predicts $\mathcal{O}(1)~{\rm event}/({\rm m^2}\times{\rm yr})$ for Planck mass DM~\cite{Bramante:2018tos}. Thus, terrestrial direct detection experiments become flux-limited. Although it is unlikely to generate the characteristic single-scatter pattern in a typical WIMP search, a heavy DM particle with a large cross-section will generate a distinctive signal of multiple scatterings in the detector~\cite{Clark:2020mna}.  Therefore, to extend the experimental sensitivity to the highest mass near the Planck scale, analyses focusing on the multiple-scattering DM are required. In this paper, we present a dedicated analysis searching for Multiply-Interacting Massive Particles~(MIMPs)~\cite{Bramante:2018qbc,Clark:2020mna} for both SI and SD interactions in the XENON1T experiment~\cite{XENON:2018voc,XENON:2019rxp}. The XENON1T experiment employs a dual-phase time projection chamber (TPC) with 2\,tonnes of ultra-pure liquid xenon in the target volume, located at INFN Laboratori Nazionali del Gran Sasso (LNGS) under 3600\,m water-equivalent overburden~\cite{XENON:2017lvq}.

It is common practice to report the cross-sections for the SI, SD-neutron, and SD-proton interactions separately~\cite{Tovey:2000mm}; although an actual MIMP, denoted as $\chi$, would possibly undergo all three interactions simultaneously~\cite{Freytsis:2010ne}. For SI scattering, the differential MIMP-nucleus scattering cross-section, $\mathrm{d}\sigma_{A,\chi}^{\mathrm{SI}}/\mathrm{d}q^2$, as a function of momentum transfer, $q$, is
\begin{equation}\label{eqn:SI}
  \frac{\mathrm{d}\sigma_{A,\chi}^{\mathrm{SI}}}{\mathrm{d}q^2} =  \frac{\mu^2_{A,\chi}}{\mu^2_{\mathrm{nucleon},\chi}}A^2\,|F_A(q)|^2\,\frac{\mathrm{d}\sigma_{\mathrm{nucleon},\chi}^{\mathrm{SI}}}{\mathrm{d}q^2},
\end{equation}
where $\mathrm{d}\sigma_{\mathrm{nucleon},\chi}^{\mathrm{SI}}/\mathrm{d}q^2$ is the differential MIMP-nucleon scattering cross-section. The Helm form factor, $F_A(q)$~\cite{Helm:1956zz,Hardy:2015boa}, is an approximation of the xenon nuclear structure functions, which is in good agreement with the calculations using large-scale shell models~\cite{Vietze:2014vsa,Fitzpatrick:2012ix,Hoferichter:2018acd}.
The squared ratio of the reduced MIMP-nucleus and MIMP-nucleon masses $\mu^2_{A,\chi}/\mu^2_{\mathrm{nucleon},\chi}$ is approximated to $A^2$ for $m_{\chi}$$\gg$$m_{A}$, which yields the $A^4$ enhancement for coherent SI scattering. However, this enhancement may not be universal at cross-sections above $10^{-32}$\,cm$^2$~\cite{Digman:2019wdm}. We thus also report a conservative result by setting $\mu^2_{A,\chi}/\mu^2_{\mathrm{nucleon},\chi} A^2$ to 1 in \cref{eqn:SI}. 

For SD scattering~\cite{Engel:1992bf}, the interaction can be described by the axial-vector–axial-vector Lagrangian. The SD interaction can be dominant if the SI interactions are either absent or strongly suppressed~\cite{Freytsis:2010ne}. The differential MIMP-nucleon cross-section, $\mathrm{d}\sigma_{A,\chi}^{\mathrm{SD}}/\mathrm{d}q^2$, can then be written as a function of the scattering cross-section, $\mathrm{d}\sigma_{n/p,\chi}^{\mathrm{SD}}/\mathrm{d}q^2$, between a MIMP and a single neutron or proton, 
\begin{equation}
  \frac{\mathrm{d}\sigma_{A,\chi}^{\mathrm{SD}}}{\mathrm{d}q^2}=\frac{4}{3}\frac{\pi}{2J+1} \frac{\mu^2_{A,\chi}}{\mu^2_{n/p,\chi}} S_A^{a_0=1,a_1=\mp 1}(q)\,\frac{\mathrm{d}\sigma_{n/p,\chi}^{\mathrm{SD}}}{\mathrm{d}q^2}, 
\end{equation}
with $\mu$ being the reduced masses, and $J$ being the initial ground-state angular momentum of the nucleus. Here, the axial-vector structure function, $S_A(q)$~\cite{Menendez:2012tm}, follows the definition in Ref.~\cite{XENON:2019rxp}.
For comparison across different target materials, experimental results are conventionally reported in the special cases of ``neutron-only" (``proton-only") coupling, where for $S_A(q)$, its isoscalar term $a_0$=$1$, and its isovector term $a_1$=$-1$~($1$). The xenon used in XENON1T contains two isotopes with nonzero nuclear spin, $^{129}$Xe (spin 1/2) and $^{131}$Xe (spin 3/2), with abundances measured to be 26.4\,\% and 21.2\,\%, respectively, with $<$1\,\% uncertainty~\cite{XENON:2019rxp}. We report our results using the $S_A(q)$ of the two xenon isotopes following the calculations by Klos et al.~\cite{Klos:2013rwa,Menendez:2012tm}, Ressell and Dean~\cite{Ressell:1997kx}, and Toivanen et al.~\cite{Toivanen:2009zza}.

As a MIMP in the mass range considered here carries high momentum, each scattering causes negligible deviation from its path, such that all recoils are aligned in a co-linear track~\cite{Bramante:2018qbc,Clark:2020mna}. We calculate the energy loss of MIMPs in the Earth's overburden before reaching the detector, by accounting for Earth's composition~\cite{Klos:2013rwa,Bednyakov:2004xq} and isotope abundances~\cite{Bramante:2018qbc}. We find the energy loss from the shielding is less than 0.1\,\% for both SI and SD cases considered in this study, with MIMP masses above 10$^{12}$\,GeV/c$^2$. Therefore, we neglect the correction on the MIMP energy and flux due to their propagation through Earth. 

 At Galactic speeds ($\sim$230\,km/s), the median MIMP passage time through the XENON1T TPC is $\mathcal{O}(1)$\,\textmu s. Thus, scintillation (S1) signals from a MIMP track will merge into a single pulse of that time scale. The data processor for XENON1T~\cite{xenon_collaboration_2018_1195785} will misidentify such a wide pulse as an ionization (S2) signal, since the merged pulse lacks the canonical single scatter S1 structure of a single maximum with a decay~\cite{XENON:2017lvq}. For clarity, we will refer to such merged pulses as an S1$_\textrm{m}$ pulse, which is a smoking-gun signature for MIMPs. After an initial S1$_\textrm{m}$ pulse, electrons freed from multiple spatially separated recoil sites will be drifted to the xenon gas to create a chain of multiple S2s. These will be reconstructed as one or more large S2 pulses, depending on the angle of the MIMP track through the TPC.

\begin{figure}
  \centering
  \includegraphics[width=0.48\textwidth]{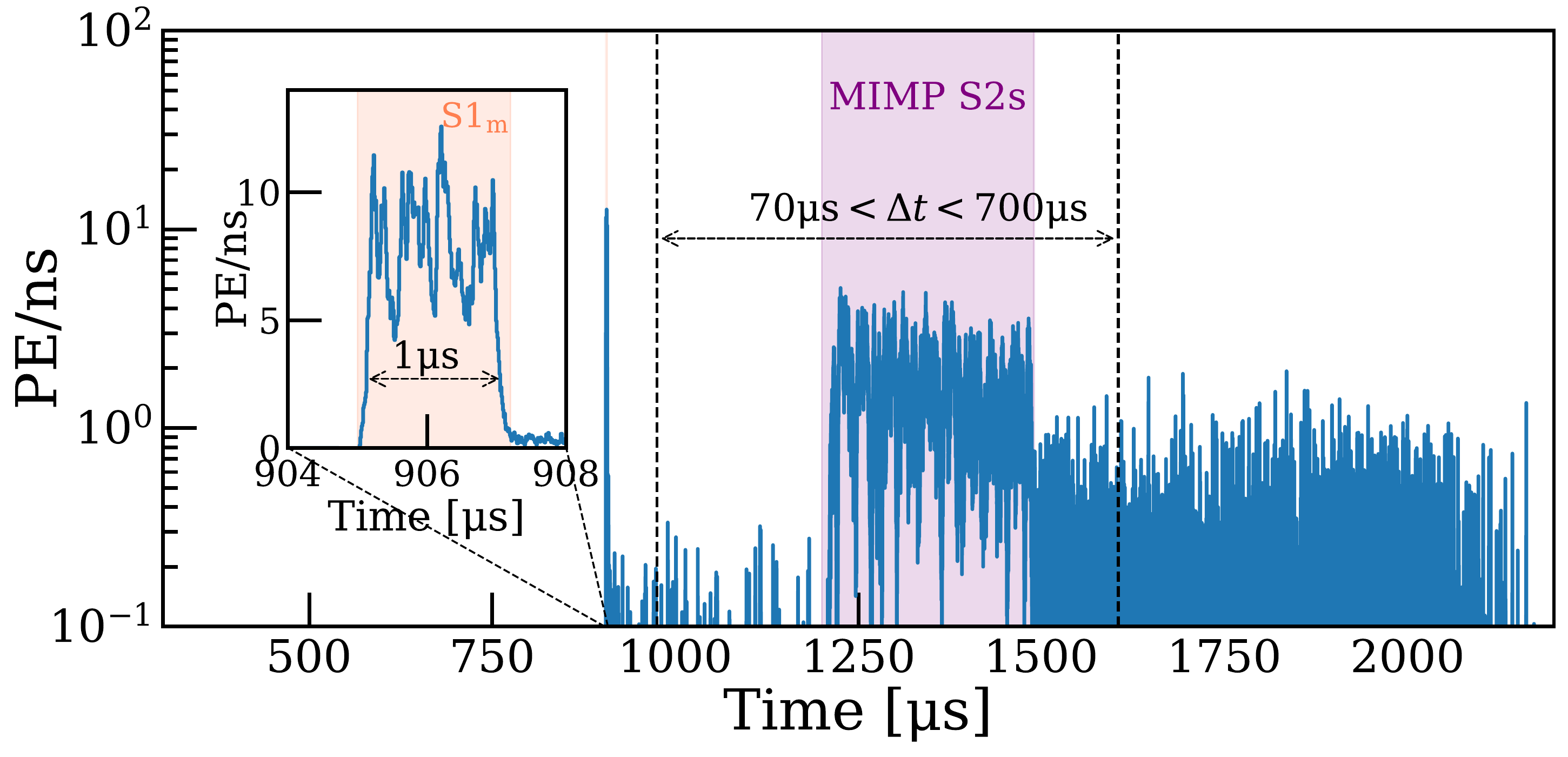}
  \caption{An example waveform of a simulated MIMP interaction in XENON1T. The simulated spin-independent scattering cross-section is $10^{-29}\,\textrm{cm}^2$, corresponding to an NR rate of  500 per meter track ($\lambda$=500\,NRs/m). The pulse in the orange band (enlarged in the insert) is the S1$_\textrm{m}$ signal. The pulses in the purple band are the S2s, which occur within the (70,700)\,\textmu s time window after the initial S1$_\textrm{m}$ signal. The pulses outside the two bands are mostly single-electron signals due to photo-ionization~\cite{XENON100:2013wdu}. }
  \label{fig:mimp_wf}
\end{figure}

As described in Ref.~\cite{Clark:2020mna}, we simulate MIMP tracks isotropically from a sphere that encloses the detector. The velocity $\vec{v}$ of the MIMP is selected with a standard isothermal DM halo velocity distribution converted to the lab frame, denoted as $f(\vec{v})$, which is parameterized as in Ref.~\cite{XENON:2018voc}. For each MIMP track across the detector, we define the per-meter nuclear recoil (NR) multiplicity as $\lambda$:
\begin{equation}
\lambda = n_{\mathrm{Xe}}\int \mathrm{d}\vec{v} \int \mathrm{d}q^2 \, \frac{\mathrm{d}\sigma_{A,\chi}}{\mathrm{d}q^2} \, f(\vec{v}),
  \label{eqn:dNdl}
\end{equation}
where $n_{\mathrm{Xe}}$ is the number density of xenon nuclei, $\mathrm{d}\sigma_{A,\chi}/\mathrm{d}q^2$ is the interaction-dependent differential cross-section using the implementation in Ref.~\cite{jelle_aalbers_2022_7041453}. The distance that a MIMP travels inside the liquid xenon before an NR occurs follows the exponential probability of $1/\lambda$. A simulated MIMP track with a random velocity drawn from $f(\vec{v})$ is first determined, followed by simulations of the vertices and time for each NR. The numbers of photons and electrons created in each recoil are modeled following Ref.~\cite{szydagis_m_2022_6808388}. The pulses detected by each photomultiplier tube (PMT) from a MIMP event are then simulated by a waveform simulator~\cite{XENON:2019izt}, which models S1s and S2s in the detector. \autoref{fig:mimp_wf} shows an example simulated waveform for a MIMP crossing the TPC. We also simulate the dominant inelastic contribution for SD scattering following Ref.~\cite{Baudis:2013bba}: immediate gamma emissions after the initial nuclear recoils ($<$1\,ns), with energies of 39.6\,keV and 80.2\,keV for $^{129}$Xe and $^{131}$Xe, respectively. Since the branching ratio of inelastic scattering is only $\sim$2\,\% for the SD interactions, we find only a percent-level change in the total acceptance of the MIMP signal. The maximum cross-section considered in this analysis is limited by the computational challenges in simulating the large number of interactions caused by a passing MIMP. Simulated tracks crossing the top 10\,cm of the 96\,cm tall TPC were excluded to ensure a minimum of 70\,\textmu s time separation between the S1$_\textrm{m}$ and the S2s for best signal identification. This leads to an average MIMP track length of 73\,cm in the TPC, and a 38\,\% decrease in the total MIMP flux. 

We used 219.4\,live-days of XENON1T data taken during 2017/2018~\cite{XENON:2018voc}, with the full 2-tonne active volume of the TPC. Multiplying the livetime by the detector's effective cross-sectional area of 0.86\,m$^2$, calculated using toy Monte Carlo, resulted in a final exposure of 188.7\,m$^2\times$day. The data was reprocessed to remove a software limit on the number of photons detected per pulse and kept blinded before all selection criteria were finalized. However, 4\,\% of the data was not blinded and was instead utilized for validation purposes in conjunction with the aforementioned simulation, for determining the MIMP selection criteria as follows.

\begin{figure}
  \centering
  \includegraphics[width=0.48\textwidth, trim=0mm 0mm 0mm -5mm, clip]{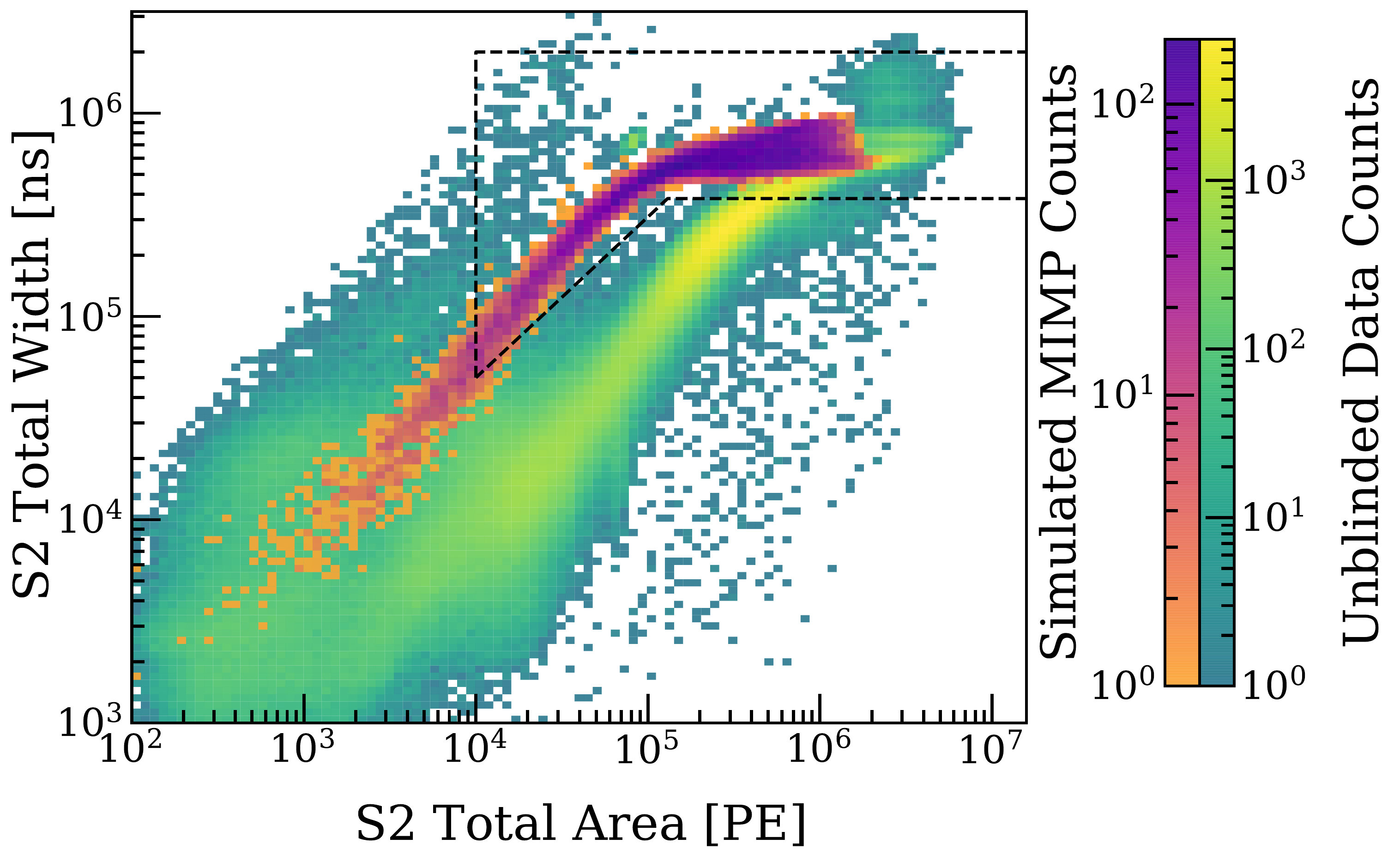} 
  \caption{The total area versus total width distribution for the 4\,\% non-blinded background events (green-yellow) and the simulated MIMP (orange-purple), after applying all other data selections mentioned in the text \textit{except} the S1$_\textrm{m}$ selection in \autoref{fig:stwid}. The selections on total area and total width are defined as the region inside the dashed lines. }
  \label{fig:TATW}
\end{figure}

The S2s contribute to the majority of the pulses in a MIMP event. To select the MIMP signal, we use the sum of the integrated pulse \textit{areas}, and the sum of the \textit{widths} (defined as the time difference between the 25\,\% to 75\,\% quantile of the pulse area) of these S2s. For the same total area, we expect MIMPs to have a higher total width compared to a single scatter electronic recoil due to their spread in time. \autoref{fig:TATW} shows the total area versus total width distributions for the non-blinded data and the simulated MIMPs. 

\begin{figure}
  \centering
  \includegraphics[width=0.48\textwidth, clip]{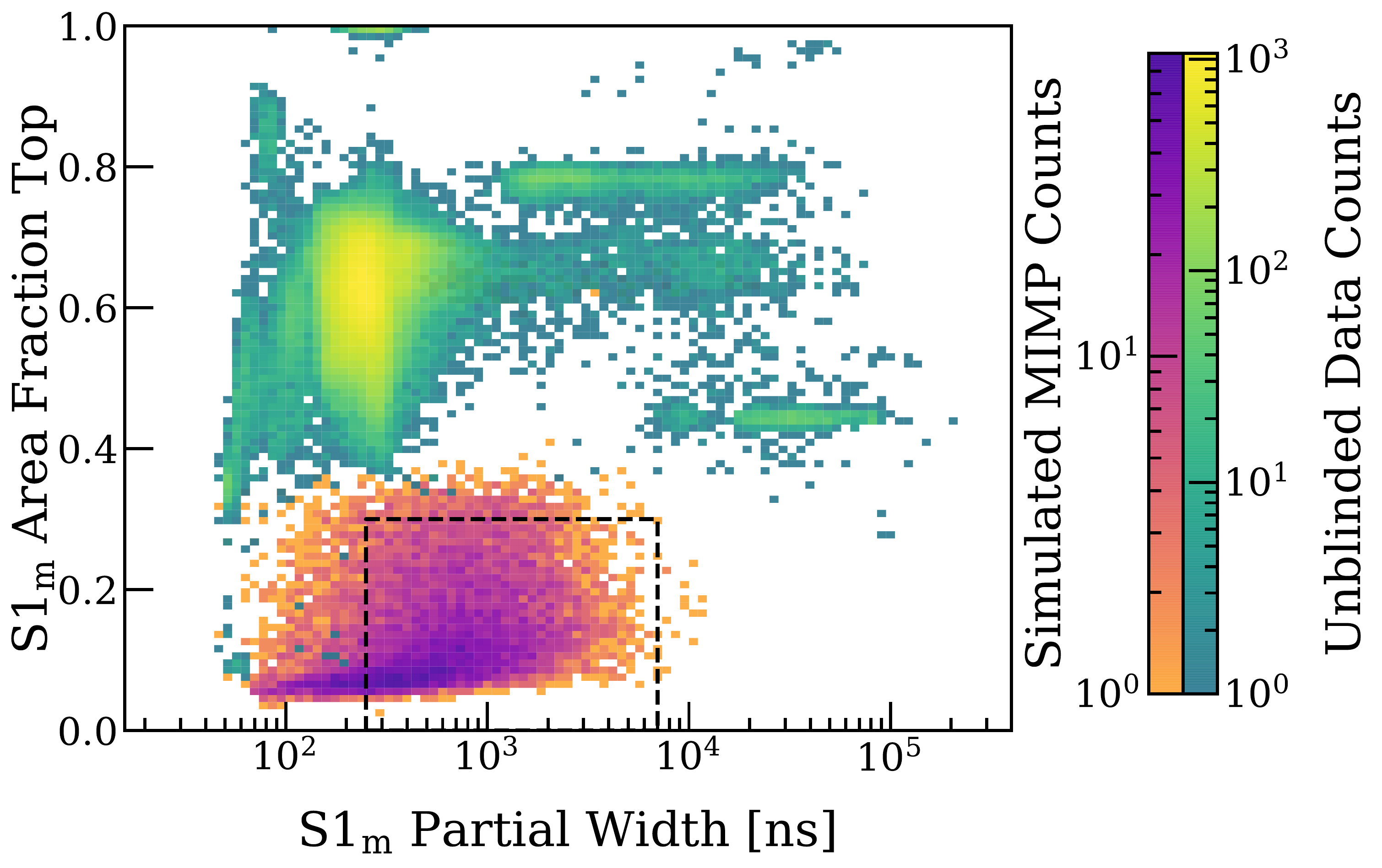} 
\caption{The partial width versus AFT distribution for the 4\,\% non-blinded background events (green-yellow) and the simulated MIMP (orange-purple), after applying all other data selections mentioned in the text. The S1$_\textrm{m}$ candidate selection is defined as the region inside the dashed square. }
  \label{fig:stwid}
\end{figure}

The S1$_\textrm{m}$ pulse is a distinctive signal for MIMPs. The \textit{Area Fraction Top}~(AFT) is defined as the fraction of pulse area seen by the top PMT array. Prompt scintillation (S1$_\textrm{m}$ or S1) from the liquid xenon gives AFT values below 0.35 due to total internal reflection on the liquid/gas interface, while proportional scintillation (S2) from the gaseous xenon gives AFT values between 0.4 and 0.8. The \textit{partial width} is defined as the time difference between the 10\,\% to 50\,\% quantile of the pulse area. The average partial width of the simulated S1$_\textrm{m}$ is 0.8\,\textmu s for a MIMP interaction, which is distinctive from the partial widths of the typical S1s ($\sim$0.2\,\textmu s) and S2s ($\sim$10\,\textmu s) in XENON1T~\cite{XENON:2017lvq}. When combined, the small AFT values and large partial widths provide distinctive signatures for MIMPs, characteristic of their non-relativistic trajectories through the TPC. We select the first peak above 250\,PE that is categorized as an S2 as the potential S1$_\textrm{m}$ candidate. To avoid the background of consecutive radioactive decays, any event with additional S1 pulses above 200\,PE is also rejected. \autoref{fig:stwid} shows the partial width versus AFT distributions of the candidate S1$_\textrm{m}$ pulses in both simulation and non-blinded data after all the selections mentioned previously, plus a muon-veto selection on data (discussed below). Only 0.04\,\% of the events in data have S1$_\textrm{m}$ candidate pulses with AFT values below 0.3, which happens when a single S1 or multiple S1s are   misidentified as an S2. We analyze the waveform patterns of those events with S1$_\textrm{m}$ partial widths above 100\,ns, and identify their origin from the decay chain of the trace amount of $^{220}$Rn within the TPC. In the decay chain, the beta-gamma decay of $^{212}$Bi is followed by the alpha decay of $^{212}$Po after a delay of approximately 200\,ns, resulting in their S1 pulses overlapping. $^{222}$Rn can produce a similar signal from the $^{214}$BiPo decays with a rarer occurrence. Since actual MIMPs can have much larger partial widths, observing all MIMP signals with partial widths $<$$250$\,ns is very unlikely ($p$-value$<$10$^{-5}$). To avoid the observed $^{212/214}$BiPo backgrounds and the potential multiple-scattering neutron background, the minimum value of the partial width parameter was chosen to be 250\,ns, and the MIMP search region is defined within the dashed lines in \autoref{fig:stwid}. 

The cosmogenic muon flux in the underground Hall~B of LNGS is about 1.2\,events$/(\mathrm{m}^2\times\mathrm{hour})$~\cite{Aprile:2014zvw,LVD:1998lir}. Unlike MIMPs, GeV muons travel at relativistic speeds, producing S1$_\textrm{m}$ pulses with shorter partial widths ($<$200\,ns), which can be vetoed by the aforementioned MIMP selections with $>$99\,\% efficiency. However, muons that travel through the top of the TPC will produce ionization signals in the gaseous xenon simultaneously with scintillation signals in the liquid. This results in a broader S1$_\textrm{m}$ pulse similar to MIMPs. By utilizing the 10-meter water Cherenkov detector surrounding the TPC~\cite{Aprile:2014zvw}, we implement the muon veto selection criteria as defined in Ref.~\cite{XENON:2019ykp}. The muon veto system achieves a 99.5\,\% muon tagging efficiency while only incurring a 1\,\% exposure loss. Prior to unblinding the full data set, we unblinded all events within 1\,\textmu s of a muon veto trigger, finding 10 muon-induced events in the MIMP signal region. This yields a background expectation of 0.05~events attributed to un-vetoed muon leakage in the total exposure.

\begin{figure}
  \centering
  \includegraphics[width=0.45\textwidth]{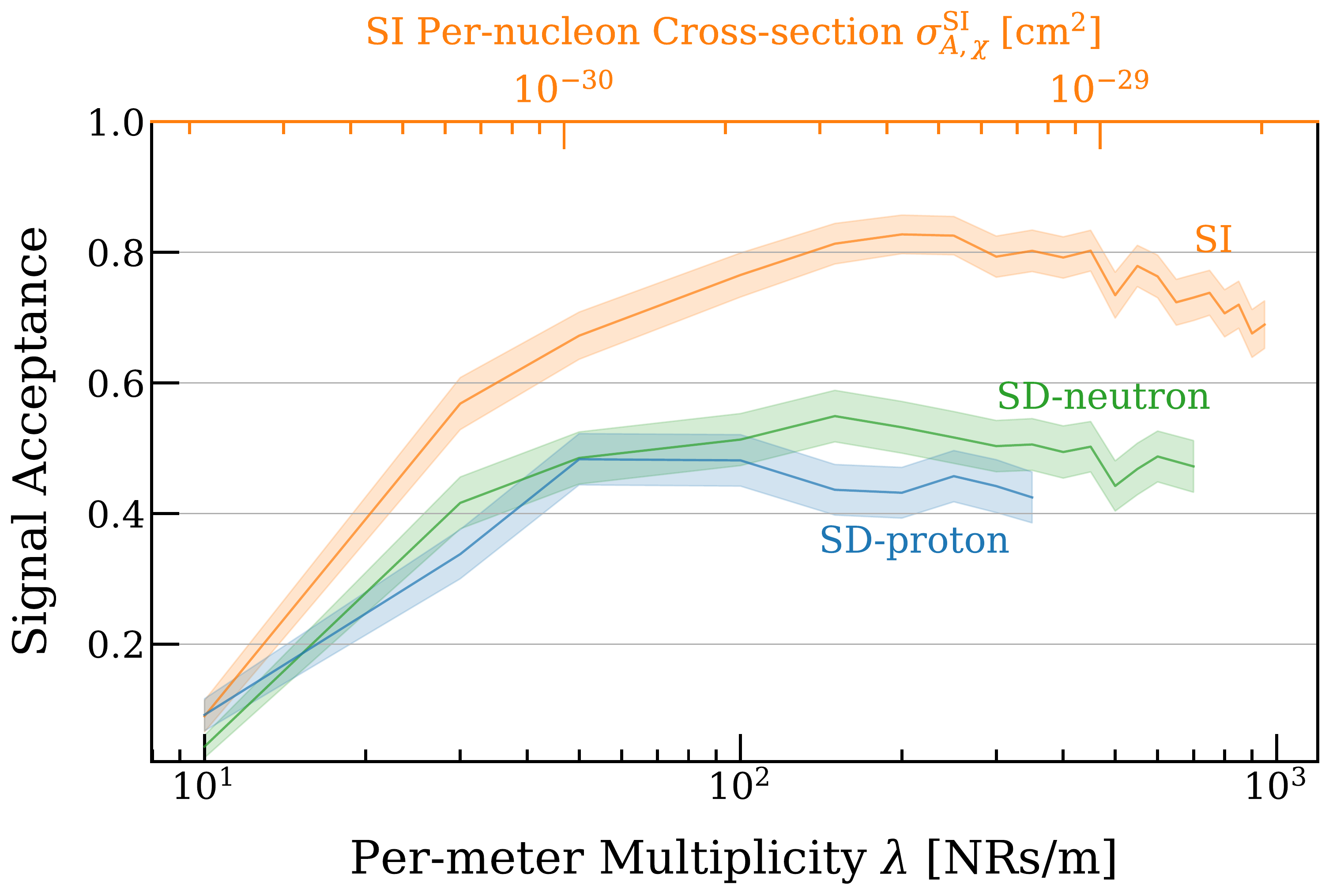} 
  \caption{The signal acceptance of MIMPs with per-meter multiplicity $\lambda$ for applied selection criteria, for SI scattering (orange), SD neutron-only scattering (green), and SD proton-only scattering (blue). The top x-axis shows the corresponding MIMP-nucleus SI cross-section assuming coherent enhancement. The uncertainty bands denote the 90\,\% Wilson confidence intervals. This acceptance does not account for the 38\,\% decrease in the MIMP flux due to the top 10\,cm veto volume. }
  \label{fig:acceptance}
\end{figure}

The probability of a MIMP signal passing our selection criteria as a function of per-meter NR multiplicity $\lambda$ is shown in \autoref{fig:acceptance}, with the corresponding SI MIMP-nucleus cross-section shown for reference. The total signal efficiency is driven by the S1$_\textrm{m}$ selection, where MIMPs with short track lengths are rejected due to their similarity with double-S1 backgrounds. The current data-processing algorithm uses \textit{Jenks optimization}~\cite{Jenks1967TheDM}, which splits a waveform into multiple pulses based on the criteria of minimizing overall deviation. When a MIMP track includes NRs with higher energies ($\gtrsim$\,70\,keV), the fluctuation in the S1$_\textrm{m}$ waveform becomes more significant. As a result, the partial width of S1$_\textrm{m}$ for MIMPs decreases due to peak-splitting, which in turn reduces the efficiency of finding the S1$_\textrm{m}$ peaks. This effect is more apparent in the SD MIMP-nucleus scattering, as their high-energy recoils occur with a higher relative frequency, leading to a decrease in their signal acceptance.

\begin{figure*}[!t]
  \centering
  \includegraphics[width=0.48\textwidth]{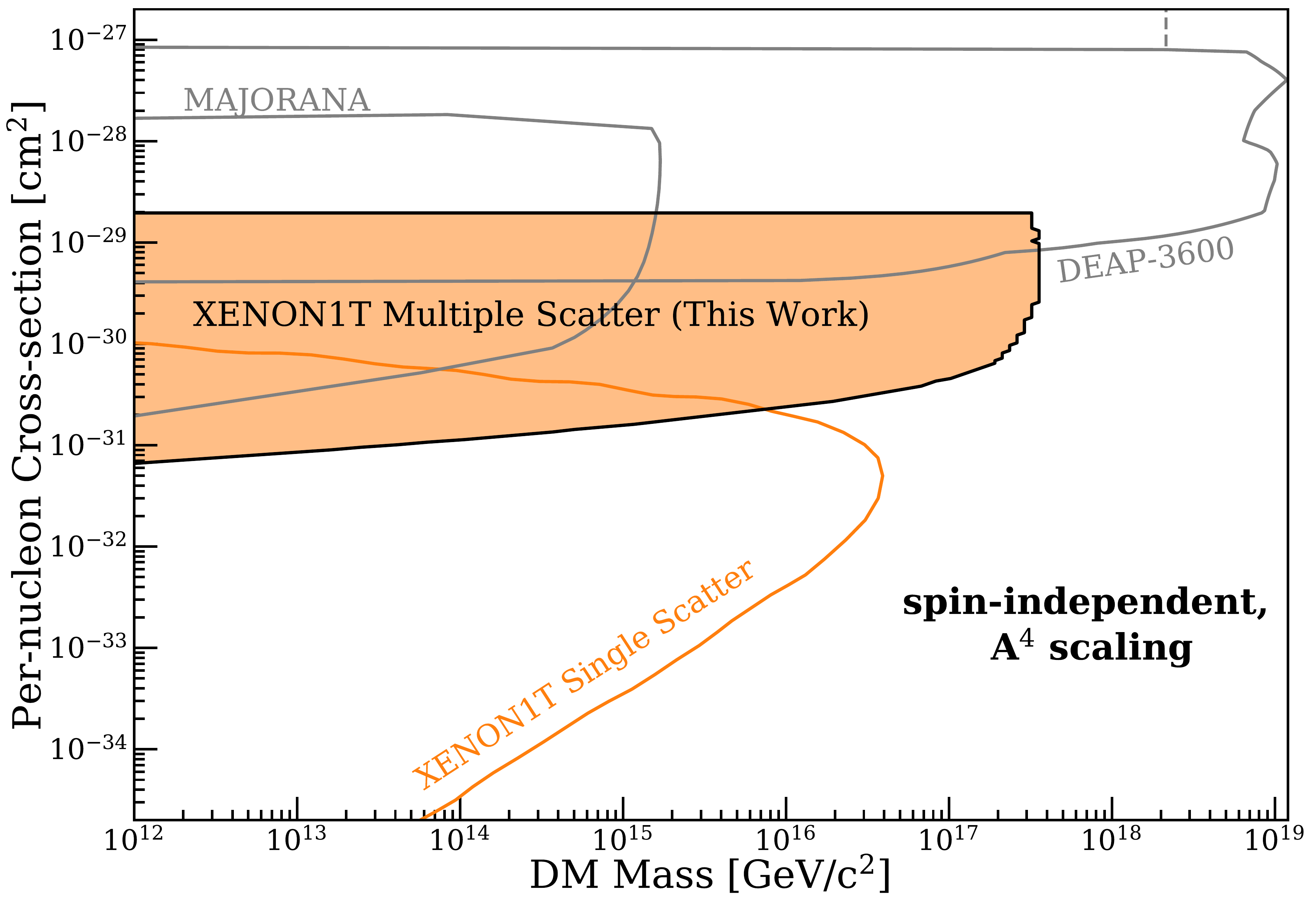}
  \includegraphics[width=0.48\textwidth]{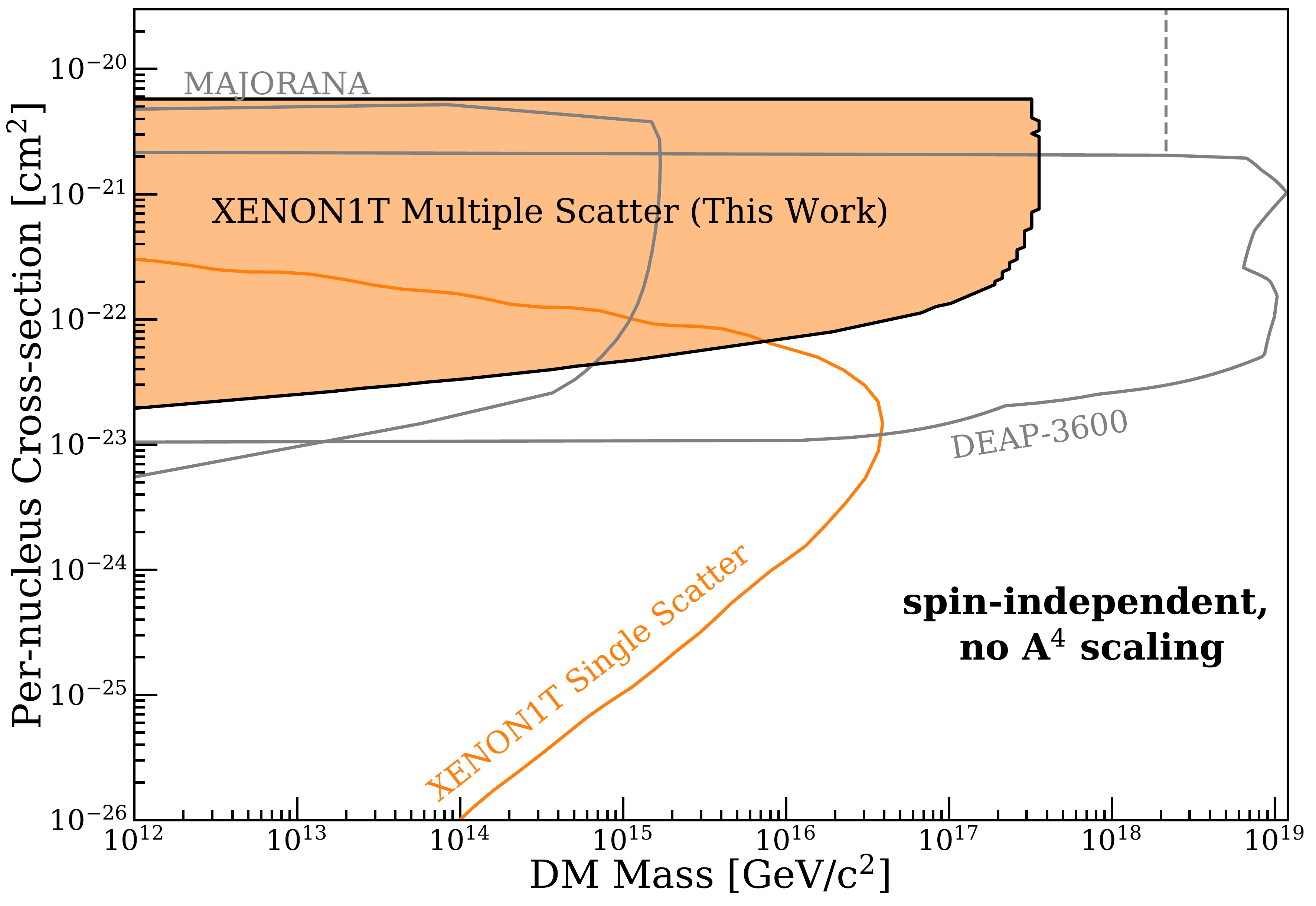} 
  \caption{The XENON1T 90\,\%~confidence level constraints on the MIMP spin-independent cross-section (orange shaded) from this multiple-scatter analysis, with (left) and without (right) the $A^4$ coherence enhancement. For comparison, we show the results from the XENON1T single-scatter analysis~\cite{Clark:2020mna} (orange line), MAJORANA Demonstrator~\cite{Majorana:2018gib,Clark:2020mna} using germanium targets and DEAP-3600~\cite{DEAPCollaboration:2021raj} using argon targets. The dashed line denotes the extrapolation region in the DEAP analysis. }
  \label{fig:sens_SI}
\end{figure*}
\begin{figure*}[!t]
  \centering
  \includegraphics[width=0.48\textwidth]{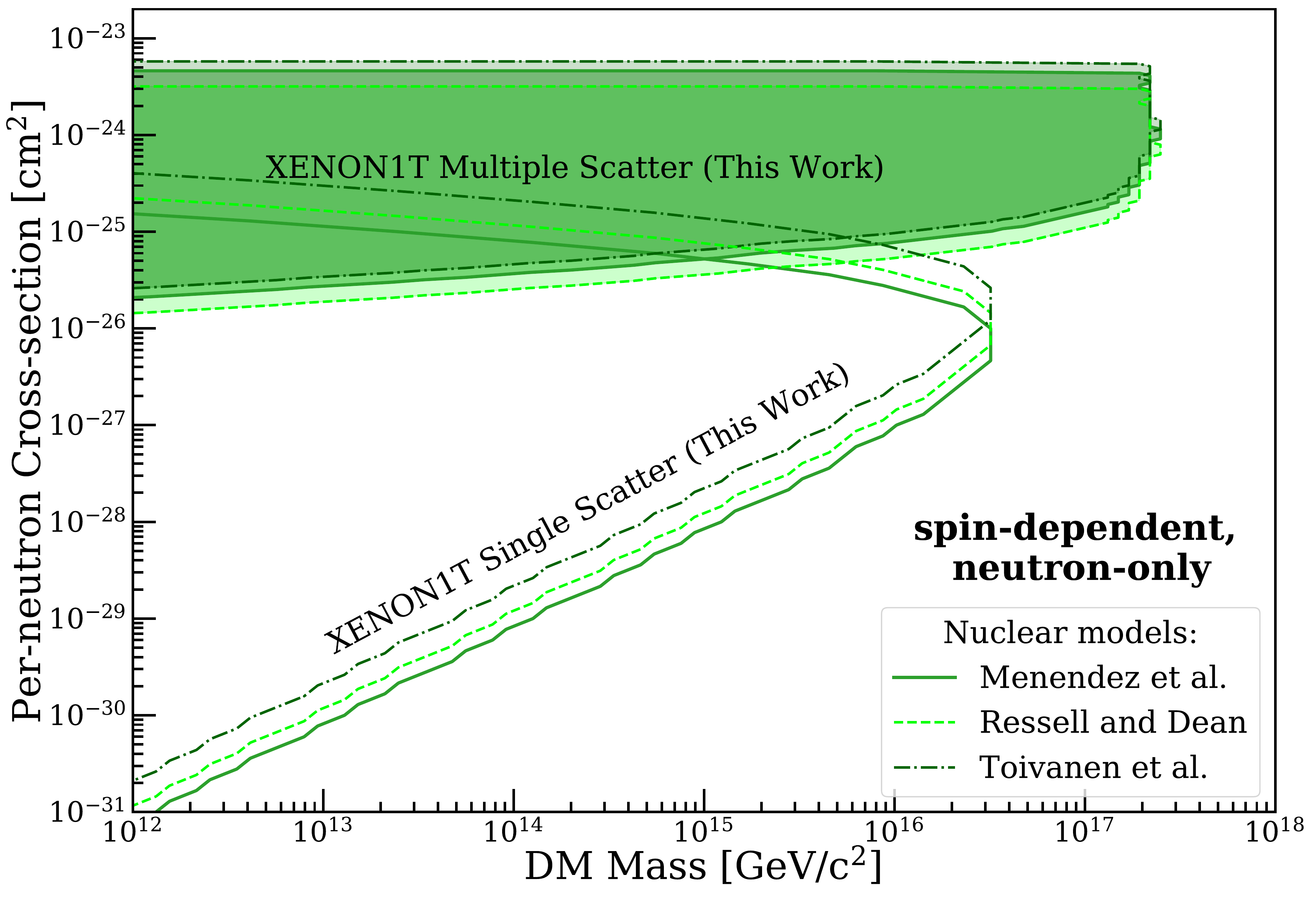}
  \includegraphics[width=0.48\textwidth]{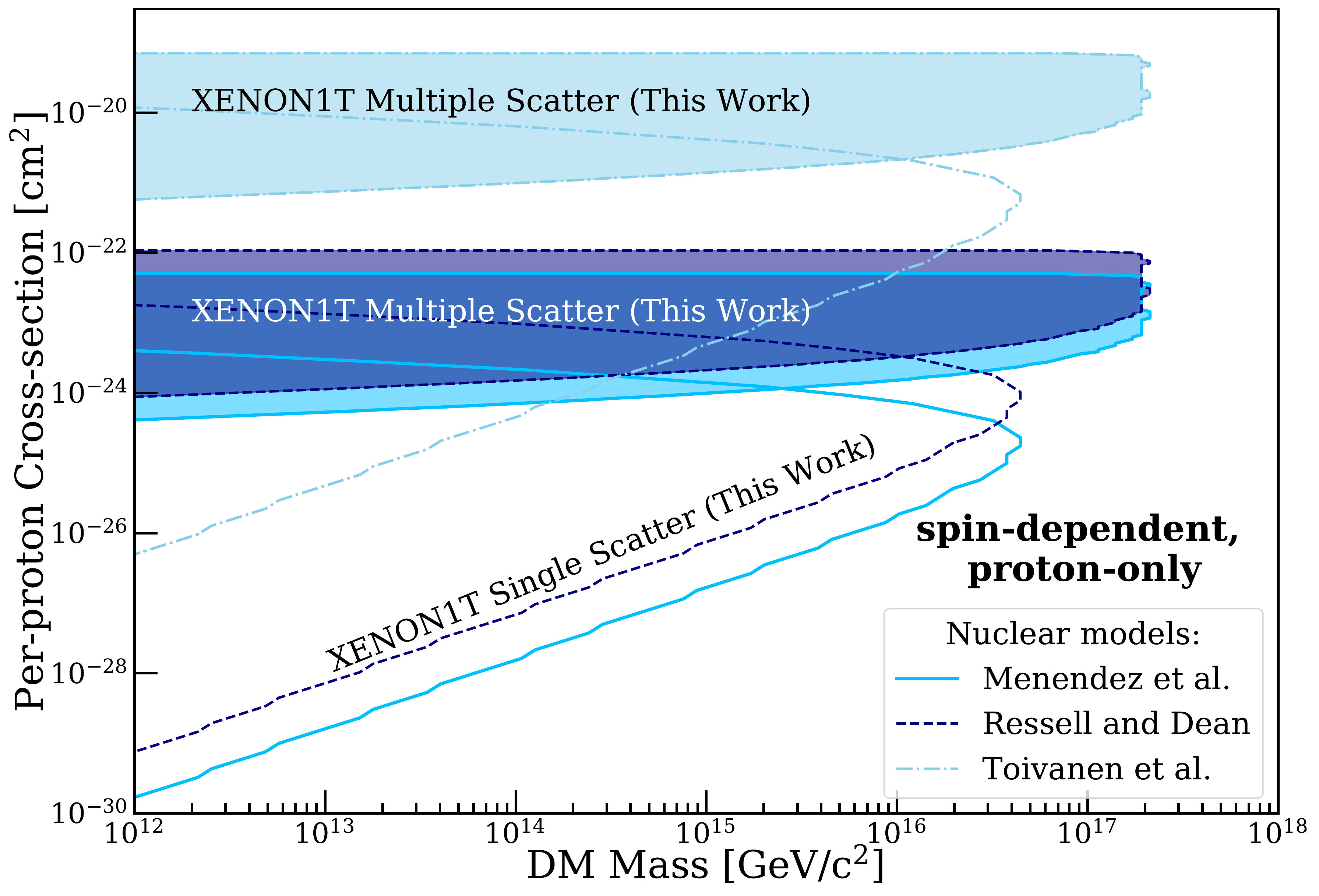} 
  \caption{The XENON1T 90\,\%~confidence level constraints on the MIMP-neutron (left, green) and MIMP-proton (right, blue) cross-sections. We show the constraints from this XENON1T multiple-scatter analysis (shaded) and a recast of the XENON1T spin-dependent single-scatter analysis~\cite{XENON:2019rxp} (line). Given different theoretical calculations of xenon nuclear models, we show our results based on the works by Klos et al.~\cite{Klos:2013rwa} (solid), Ressell and Dean~\cite{Ressell:1997kx} (dashed), and Toivanen et al.~\cite{Toivanen:2009zza} (dash-dotted).}
  \label{fig:sens_SD}
\end{figure*}
Upon unblinding, no MIMP candidate events were found that would satisfy the selections. This is in agreement with our total background expectation of 0.05~events in the full exposure. Hence, we derive MIMP exclusion limits following the Feldman-Cousins~\cite{Feldman:1997qc} procedure. The calculation uses the aforementioned MIMP flux and interaction models, and the detection efficiency in \autoref{fig:acceptance}. We show in \autoref{fig:sens_SI} the 90\,\% confidence level constraints on the SI MIMP-nucleon cross-section for masses above $10^{12}$\,GeV/c$^2$, with and without the assumed $A^4$ enhancement for coherent scattering. Compared to the standard single-scatter search, this analysis extends the mass reach of XENON1T by one order of magnitude to 3$\times$10$^{17}$\,GeV/c$^2$. In the case of $A^4$ enhancement, the limit covers new parameter space spanning an order of magnitude in cross-section and three orders of magnitude in mass. 
We show in \autoref{fig:sens_SD} the 90\,\% confidence level constraints on the SD MIMP-nucleon cross-section for masses above $10^{12}$\,GeV/c$^2$, assuming neutron- or proton-only coupling. We report these results based on the calculations done in Ref.~\cite{Klos:2013rwa} using the median of their structure function, as well as using alternative nuclear models~\cite{Ressell:1997kx,Toivanen:2009zza}. This analysis rules out approximately two orders of magnitude in MIMP-neutron (MIMP-proton) cross-section, up to a mass of 2$\times$10$^{17}$\,GeV/c$^2$. The previous XENON1T single-scatter WIMP results~\cite{XENON:2019rxp} using a 0.65\,tonne fiducial mass are recast following the SD MIMP model. Other SD exclusion limits exist but are not shown here: DAMA reported a MIMP-proton limit in 1999~\cite{Bernabei:1999ui}, but assumed a structure function of unity; a recent MIMP-proton limit from PICO-60 is derived in a Ph.D. thesis~\cite{Broerman:2022jtj}. 

In summary, we conducted a blind search for track-like signals from heavy, multiply scattering DM using 219.4~days of data from XENON1T. Following unblinding, we found no signal candidate events and thus calculated exclusion regions for both spin-independent and spin-dependent DM reaching masses up to 3$\times$10$^{17}$\,GeV/c$^2$ and 2$\times$10$^{17}$\,GeV/c$^2$, respectively. These are the first multiple-scatter constraints set by a xenon-based experiment, and the first spin-dependent limits for MIMP-neutron and MIMP-proton interactions. 
Further sensitivity at even higher cross-sections may come from searches using mica slabs~\cite{Acevedo:2021tbl} or etched plastic~\cite{Bhoonah:2020fys}. As the DM flux becomes the limiting factor, higher MIMP masses will soon be probed by other liquid noble element detectors which have a larger exposure (measured in area$\times$time). 
XENONnT~\cite{XENON:2020kmp} and LZ~\cite{LZ:2018qzl} are running and may extend this sensitivity by an order of magnitude in mass, and the upcoming DarkSide-20k~\cite{DarkSide-20k:2017zyg} and DARWIN/XLZD~\cite{DARWIN:2016hyl, Aalbers:2022dzr} detectors may push this sensitivity beyond the Planck mass. Thus, even Planck-scale physics is within reach of direct DM experiments.

\begin{acknowledgments}
We thank Javier Menendez, Achim Schwenk, and Martin Hoferichter for the fruitful discussions, and Javier Menendez for the calculation of the spin expectation values for the $^{25}$Mg isotope used for Earth's shielding effect. We thank Hayden Schennum for his contribution to the track simulation. We gratefully acknowledge support from the National Science Foundation, Swiss National Science Foundation, German Ministry for Education and Research, Max Planck Gesellschaft, Deutsche Forschungsgemeinschaft, Helmholtz Association, Dutch Research Council (NWO), Weizmann Institute of Science, Israeli Science Foundation, Binational Science Foundation, Fundacao para a Ciencia e a Tecnologia, R\'egion des Pays de la Loire, Knut and Alice Wallenberg Foundation, Kavli Foundation, JSPS Kakenhi and JST FOREST Program in Japan, Tsinghua University Initiative Scientific Research Program and Istituto Nazionale di Fisica Nucleare. This project has received funding/support from the European Union’s Horizon 2020 research and innovation programme under the Marie Sk\l{}odowska-Curie grant agreement No 860881-HIDDeN. Data processing is performed using infrastructures from the Open Science Grid, the European Grid Initiative and the Dutch national e-infrastructure with the support of SURF Cooperative. We are grateful to Laboratori Nazionali del Gran Sasso for hosting and supporting the XENON project.
\end{acknowledgments}

\bibliography{ref}

\begin{thebibliography}{60}%
\makeatletter
\providecommand \@ifxundefined [1]{%
 \@ifx{#1\undefined}
}%
\providecommand \@ifnum [1]{%
 \ifnum #1\expandafter \@firstoftwo
 \else \expandafter \@secondoftwo
 \fi
}%
\providecommand \@ifx [1]{%
 \ifx #1\expandafter \@firstoftwo
 \else \expandafter \@secondoftwo
 \fi
}%
\providecommand \natexlab [1]{#1}%
\providecommand \enquote  [1]{``#1''}%
\providecommand \bibnamefont  [1]{#1}%
\providecommand \bibfnamefont [1]{#1}%
\providecommand \citenamefont [1]{#1}%
\providecommand \href@noop [0]{\@secondoftwo}%
\providecommand \href [0]{\begingroup \@sanitize@url \@href}%
\providecommand \@href[1]{\@@startlink{#1}\@@href}%
\providecommand \@@href[1]{\endgroup#1\@@endlink}%
\providecommand \@sanitize@url [0]{\catcode `\\12\catcode `\$12\catcode
  `\&12\catcode `\#12\catcode `\^12\catcode `\_12\catcode `\%12\relax}%
\providecommand \@@startlink[1]{}%
\providecommand \@@endlink[0]{}%
\providecommand \url  [0]{\begingroup\@sanitize@url \@url }%
\providecommand \@url [1]{\endgroup\@href {#1}{\urlprefix }}%
\providecommand \urlprefix  [0]{URL }%
\providecommand \Eprint [0]{\href }%
\providecommand \doibase [0]{http://dx.doi.org/}%
\providecommand \selectlanguage [0]{\@gobble}%
\providecommand \bibinfo  [0]{\@secondoftwo}%
\providecommand \bibfield  [0]{\@secondoftwo}%
\providecommand \translation [1]{[#1]}%
\providecommand \BibitemOpen [0]{}%
\providecommand \bibitemStop [0]{}%
\providecommand \bibitemNoStop [0]{.\EOS\space}%
\providecommand \EOS [0]{\spacefactor3000\relax}%
\providecommand \BibitemShut  [1]{\csname bibitem#1\endcsname}%
\let\auto@bib@innerbib\@empty
\bibitem [{\citenamefont {Billard}\ \emph {et~al.}(2022)\citenamefont {Billard}
  \emph {et~al.}}]{Billard:2021uyg}%
  \BibitemOpen
  \bibfield  {author} {\bibinfo {author} {\bibfnamefont {J.}~\bibnamefont
  {Billard}} \emph {et~al.},\ }\href {\doibase 10.1088/1361-6633/ac5754}
  {\bibfield  {journal} {\bibinfo  {journal} {Rept. Prog. Phys.}\ }\textbf
  {\bibinfo {volume} {85}},\ \bibinfo {pages} {056201} (\bibinfo {year}
  {2022})},\ \Eprint {http://arxiv.org/abs/2104.07634} {arXiv:2104.07634
  [hep-ex]} \BibitemShut {NoStop}%
\bibitem [{\citenamefont {Chu}\ \emph {et~al.}(2012)\citenamefont {Chu},
  \citenamefont {Hambye},\ and\ \citenamefont {Tytgat}}]{Chu:2011be}%
  \BibitemOpen
  \bibfield  {author} {\bibinfo {author} {\bibfnamefont {X.}~\bibnamefont
  {Chu}}, \bibinfo {author} {\bibfnamefont {T.}~\bibnamefont {Hambye}}, \ and\
  \bibinfo {author} {\bibfnamefont {M.~H.~G.}\ \bibnamefont {Tytgat}},\ }\href
  {\doibase 10.1088/1475-7516/2012/05/034} {\bibfield  {journal} {\bibinfo
  {journal} {JCAP}\ }\textbf {\bibinfo {volume} {05}},\ \bibinfo {pages} {034}
  (\bibinfo {year} {2012})},\ \Eprint {http://arxiv.org/abs/1112.0493}
  {arXiv:1112.0493} \BibitemShut {NoStop}%
\bibitem [{\citenamefont {Berlin}\ \emph {et~al.}(2016)\citenamefont {Berlin},
  \citenamefont {Hooper},\ and\ \citenamefont {Krnjaic}}]{Berlin:2016gtr}%
  \BibitemOpen
  \bibfield  {author} {\bibinfo {author} {\bibfnamefont {A.}~\bibnamefont
  {Berlin}}, \bibinfo {author} {\bibfnamefont {D.}~\bibnamefont {Hooper}}, \
  and\ \bibinfo {author} {\bibfnamefont {G.}~\bibnamefont {Krnjaic}},\ }\href
  {\doibase 10.1103/PhysRevD.94.095019} {\bibfield  {journal} {\bibinfo
  {journal} {Phys. Rev. D}\ }\textbf {\bibinfo {volume} {94}},\ \bibinfo
  {pages} {095019} (\bibinfo {year} {2016})},\ \Eprint
  {http://arxiv.org/abs/1609.02555} {arXiv:1609.02555} \BibitemShut {NoStop}%
\bibitem [{\citenamefont {Chung}\ \emph {et~al.}(1999)\citenamefont {Chung},
  \citenamefont {Kolb},\ and\ \citenamefont {Riotto}}]{Chung:1998rq}%
  \BibitemOpen
  \bibfield  {author} {\bibinfo {author} {\bibfnamefont {D.~J.~H.}\
  \bibnamefont {Chung}}, \bibinfo {author} {\bibfnamefont {E.~W.}\ \bibnamefont
  {Kolb}}, \ and\ \bibinfo {author} {\bibfnamefont {A.}~\bibnamefont
  {Riotto}},\ }\href {\doibase 10.1103/PhysRevD.60.063504} {\bibfield
  {journal} {\bibinfo  {journal} {Phys. Rev. D}\ }\textbf {\bibinfo {volume}
  {60}},\ \bibinfo {pages} {063504} (\bibinfo {year} {1999})},\ \Eprint
  {http://arxiv.org/abs/hep-ph/9809453} {arXiv:hep-ph/9809453} \BibitemShut
  {NoStop}%
\bibitem [{\citenamefont {Giudice}\ \emph {et~al.}(2001)\citenamefont
  {Giudice}, \citenamefont {Kolb},\ and\ \citenamefont
  {Riotto}}]{Giudice:2000ex}%
  \BibitemOpen
  \bibfield  {author} {\bibinfo {author} {\bibfnamefont {G.~F.}\ \bibnamefont
  {Giudice}}, \bibinfo {author} {\bibfnamefont {E.~W.}\ \bibnamefont {Kolb}}, \
  and\ \bibinfo {author} {\bibfnamefont {A.}~\bibnamefont {Riotto}},\ }\href
  {\doibase 10.1103/PhysRevD.64.023508} {\bibfield  {journal} {\bibinfo
  {journal} {Phys. Rev. D}\ }\textbf {\bibinfo {volume} {64}},\ \bibinfo
  {pages} {023508} (\bibinfo {year} {2001})},\ \Eprint
  {http://arxiv.org/abs/hep-ph/0005123} {arXiv:hep-ph/0005123} \BibitemShut
  {NoStop}%
\bibitem [{\citenamefont {Azatov}\ \emph {et~al.}(2021)\citenamefont {Azatov},
  \citenamefont {Vanvlasselaer},\ and\ \citenamefont {Yin}}]{Azatov:2021ifm}%
  \BibitemOpen
  \bibfield  {author} {\bibinfo {author} {\bibfnamefont {A.}~\bibnamefont
  {Azatov}}, \bibinfo {author} {\bibfnamefont {M.}~\bibnamefont
  {Vanvlasselaer}}, \ and\ \bibinfo {author} {\bibfnamefont {W.}~\bibnamefont
  {Yin}},\ }\href {\doibase 10.1007/JHEP03(2021)288} {\bibfield  {journal}
  {\bibinfo  {journal} {JHEP}\ }\textbf {\bibinfo {volume} {03}},\ \bibinfo
  {pages} {288} (\bibinfo {year} {2021})},\ \Eprint
  {http://arxiv.org/abs/2101.05721} {arXiv:2101.05721} \BibitemShut {NoStop}%
\bibitem [{\citenamefont {Baldes}\ \emph {et~al.}(2022)\citenamefont {Baldes},
  \citenamefont {Gouttenoire}, \citenamefont {Sala},\ and\ \citenamefont
  {Servant}}]{Baldes:2021aph}%
  \BibitemOpen
  \bibfield  {author} {\bibinfo {author} {\bibfnamefont {I.}~\bibnamefont
  {Baldes}}, \bibinfo {author} {\bibfnamefont {Y.}~\bibnamefont {Gouttenoire}},
  \bibinfo {author} {\bibfnamefont {F.}~\bibnamefont {Sala}}, \ and\ \bibinfo
  {author} {\bibfnamefont {G.}~\bibnamefont {Servant}},\ }\href@noop {}
  {\bibfield  {journal} {\bibinfo  {journal} {Journal of High Energy Physics}\
  }\textbf {\bibinfo {volume} {2022}},\ \bibinfo {pages} {1} (\bibinfo {year}
  {2022})},\ \Eprint {http://arxiv.org/abs/2110.13926} {arXiv:2110.13926}
  \BibitemShut {NoStop}%
\bibitem [{\citenamefont {Takahashi}(2008)}]{Takahashi:2007tz}%
  \BibitemOpen
  \bibfield  {author} {\bibinfo {author} {\bibfnamefont {F.}~\bibnamefont
  {Takahashi}},\ }\href {\doibase 10.1016/j.physletb.2007.12.048} {\bibfield
  {journal} {\bibinfo  {journal} {Phys. Lett. B}\ }\textbf {\bibinfo {volume}
  {660}},\ \bibinfo {pages} {100} (\bibinfo {year} {2008})},\ \Eprint
  {http://arxiv.org/abs/0705.0579} {arXiv:0705.0579} \BibitemShut {NoStop}%
\bibitem [{\citenamefont {Acharya}\ \emph {et~al.}(2009)\citenamefont
  {Acharya}, \citenamefont {Kane}, \citenamefont {Watson},\ and\ \citenamefont
  {Kumar}}]{Acharya:2009zt}%
  \BibitemOpen
  \bibfield  {author} {\bibinfo {author} {\bibfnamefont {B.~S.}\ \bibnamefont
  {Acharya}}, \bibinfo {author} {\bibfnamefont {G.}~\bibnamefont {Kane}},
  \bibinfo {author} {\bibfnamefont {S.}~\bibnamefont {Watson}}, \ and\ \bibinfo
  {author} {\bibfnamefont {P.}~\bibnamefont {Kumar}},\ }\href {\doibase
  10.1103/PhysRevD.80.083529} {\bibfield  {journal} {\bibinfo  {journal} {Phys.
  Rev. D}\ }\textbf {\bibinfo {volume} {80}},\ \bibinfo {pages} {083529}
  (\bibinfo {year} {2009})},\ \Eprint {http://arxiv.org/abs/0908.2430}
  {arXiv:0908.2430} \BibitemShut {NoStop}%
\bibitem [{\citenamefont {Chung}\ \emph {et~al.}(1998)\citenamefont {Chung},
  \citenamefont {Kolb},\ and\ \citenamefont {Riotto}}]{Chung:1998zb}%
  \BibitemOpen
  \bibfield  {author} {\bibinfo {author} {\bibfnamefont {D.~J.~H.}\
  \bibnamefont {Chung}}, \bibinfo {author} {\bibfnamefont {E.~W.}\ \bibnamefont
  {Kolb}}, \ and\ \bibinfo {author} {\bibfnamefont {A.}~\bibnamefont
  {Riotto}},\ }\href {\doibase 10.1103/PhysRevD.59.023501} {\bibfield
  {journal} {\bibinfo  {journal} {Phys. Rev. D}\ }\textbf {\bibinfo {volume}
  {59}},\ \bibinfo {pages} {023501} (\bibinfo {year} {1998})},\ \Eprint
  {http://arxiv.org/abs/hep-ph/9802238} {arXiv:hep-ph/9802238} \BibitemShut
  {NoStop}%
\bibitem [{\citenamefont {Ema}\ \emph {et~al.}(2019)\citenamefont {Ema},
  \citenamefont {Nakayama},\ and\ \citenamefont {Tang}}]{Ema:2019yrd}%
  \BibitemOpen
  \bibfield  {author} {\bibinfo {author} {\bibfnamefont {Y.}~\bibnamefont
  {Ema}}, \bibinfo {author} {\bibfnamefont {K.}~\bibnamefont {Nakayama}}, \
  and\ \bibinfo {author} {\bibfnamefont {Y.}~\bibnamefont {Tang}},\ }\href
  {\doibase 10.1007/JHEP07(2019)060} {\bibfield  {journal} {\bibinfo  {journal}
  {JHEP}\ }\textbf {\bibinfo {volume} {07}},\ \bibinfo {pages} {060} (\bibinfo
  {year} {2019})},\ \Eprint {http://arxiv.org/abs/1903.10973}
  {arXiv:1903.10973} \BibitemShut {NoStop}%
\bibitem [{\citenamefont {MacGibbon}(1987)}]{MacGibbon:1987my}%
  \BibitemOpen
  \bibfield  {author} {\bibinfo {author} {\bibfnamefont {J.~H.}\ \bibnamefont
  {MacGibbon}},\ }\href {\doibase 10.1038/329308a0} {\bibfield  {journal}
  {\bibinfo  {journal} {Nature}\ }\textbf {\bibinfo {volume} {329}},\ \bibinfo
  {pages} {308} (\bibinfo {year} {1987})}\BibitemShut {NoStop}%
\bibitem [{\citenamefont {Maldacena}(2021)}]{Maldacena:2020skw}%
  \BibitemOpen
  \bibfield  {author} {\bibinfo {author} {\bibfnamefont {J.}~\bibnamefont
  {Maldacena}},\ }\href {\doibase 10.1007/JHEP04(2021)079} {\bibfield
  {journal} {\bibinfo  {journal} {JHEP}\ }\textbf {\bibinfo {volume} {04}},\
  \bibinfo {pages} {079} (\bibinfo {year} {2021})},\ \Eprint
  {http://arxiv.org/abs/2004.06084} {arXiv:2004.06084} \BibitemShut {NoStop}%
\bibitem [{\citenamefont {Carney}\ \emph {et~al.}(2022)\citenamefont {Carney},
  \citenamefont {Raj}, \citenamefont {Bai}, \citenamefont {Berger},
  \citenamefont {Blanco}, \citenamefont {Bramante}, \citenamefont {Cappiello},
  \citenamefont {Dutra}, \citenamefont {Ebadi}, \citenamefont {Engel} \emph
  {et~al.}}]{Carney:2022gse}%
  \BibitemOpen
  \bibfield  {author} {\bibinfo {author} {\bibfnamefont {D.}~\bibnamefont
  {Carney}}, \bibinfo {author} {\bibfnamefont {N.}~\bibnamefont {Raj}},
  \bibinfo {author} {\bibfnamefont {Y.}~\bibnamefont {Bai}}, \bibinfo {author}
  {\bibfnamefont {J.}~\bibnamefont {Berger}}, \bibinfo {author} {\bibfnamefont
  {C.}~\bibnamefont {Blanco}}, \bibinfo {author} {\bibfnamefont
  {J.}~\bibnamefont {Bramante}}, \bibinfo {author} {\bibfnamefont
  {C.}~\bibnamefont {Cappiello}}, \bibinfo {author} {\bibfnamefont
  {M.}~\bibnamefont {Dutra}}, \bibinfo {author} {\bibfnamefont
  {R.}~\bibnamefont {Ebadi}}, \bibinfo {author} {\bibfnamefont
  {K.}~\bibnamefont {Engel}},  \emph {et~al.},\ }\href@noop {} {\bibfield
  {journal} {\bibinfo  {journal} {arXiv preprint arXiv:2203.06508}\ } (\bibinfo
  {year} {2022})}\BibitemShut {NoStop}%
\bibitem [{\citenamefont {Goodman}\ and\ \citenamefont
  {Witten}(1985)}]{Goodman:1984dc}%
  \BibitemOpen
  \bibfield  {author} {\bibinfo {author} {\bibfnamefont {M.~W.}\ \bibnamefont
  {Goodman}}\ and\ \bibinfo {author} {\bibfnamefont {E.}~\bibnamefont
  {Witten}},\ }\href {\doibase 10.1103/PhysRevD.31.3059} {\bibfield  {journal}
  {\bibinfo  {journal} {Phys. Rev. D}\ }\textbf {\bibinfo {volume} {31}},\
  \bibinfo {pages} {3059} (\bibinfo {year} {1985})}\BibitemShut {NoStop}%
\bibitem [{\citenamefont {Digman}\ \emph {et~al.}(2019)\citenamefont {Digman},
  \citenamefont {Cappiello}, \citenamefont {Beacom}, \citenamefont {Hirata},\
  and\ \citenamefont {Peter}}]{Digman:2019wdm}%
  \BibitemOpen
  \bibfield  {author} {\bibinfo {author} {\bibfnamefont {M.~C.}\ \bibnamefont
  {Digman}}, \bibinfo {author} {\bibfnamefont {C.~V.}\ \bibnamefont
  {Cappiello}}, \bibinfo {author} {\bibfnamefont {J.~F.}\ \bibnamefont
  {Beacom}}, \bibinfo {author} {\bibfnamefont {C.~M.}\ \bibnamefont {Hirata}},
  \ and\ \bibinfo {author} {\bibfnamefont {A.~H.}\ \bibnamefont {Peter}},\
  }\href {\doibase 10.1103/PhysRevD.100.063013} {\bibfield  {journal} {\bibinfo
   {journal} {Phys. Rev. D}\ }\textbf {\bibinfo {volume} {100}},\ \bibinfo
  {pages} {063013} (\bibinfo {year} {2019})},\ \Eprint
  {http://arxiv.org/abs/1907.10618} {arXiv:1907.10618} \BibitemShut {NoStop}%
\bibitem [{\citenamefont {Bernabei}\ \emph {et~al.}(1999)\citenamefont
  {Bernabei} \emph {et~al.}}]{Bernabei:1999ui}%
  \BibitemOpen
  \bibfield  {author} {\bibinfo {author} {\bibfnamefont {R.}~\bibnamefont
  {Bernabei}} \emph {et~al.},\ }\href {\doibase 10.1103/PhysRevLett.83.4918}
  {\bibfield  {journal} {\bibinfo  {journal} {Phys. Rev. Lett.}\ }\textbf
  {\bibinfo {volume} {83}},\ \bibinfo {pages} {4918} (\bibinfo {year}
  {1999})}\BibitemShut {NoStop}%
\bibitem [{\citenamefont {Jungman}\ \emph {et~al.}(1996)\citenamefont
  {Jungman}, \citenamefont {Kamionkowski},\ and\ \citenamefont
  {Griest}}]{Jungman:1995df}%
  \BibitemOpen
  \bibfield  {author} {\bibinfo {author} {\bibfnamefont {G.}~\bibnamefont
  {Jungman}}, \bibinfo {author} {\bibfnamefont {M.}~\bibnamefont
  {Kamionkowski}}, \ and\ \bibinfo {author} {\bibfnamefont {K.}~\bibnamefont
  {Griest}},\ }\href {\doibase 10.1016/0370-1573(95)00058-5} {\bibfield
  {journal} {\bibinfo  {journal} {Phys. Rept.}\ }\textbf {\bibinfo {volume}
  {267}},\ \bibinfo {pages} {195} (\bibinfo {year} {1996})},\ \Eprint
  {http://arxiv.org/abs/hep-ph/9506380} {arXiv:hep-ph/9506380} \BibitemShut
  {NoStop}%
\bibitem [{\citenamefont {Aprile}\ \emph
  {et~al.}(2019{\natexlab{a}})\citenamefont {Aprile} \emph
  {et~al.}}]{XENON:2019rxp}%
  \BibitemOpen
  \bibfield  {author} {\bibinfo {author} {\bibfnamefont {E.}~\bibnamefont
  {Aprile}} \emph {et~al.} (\bibinfo {collaboration} {XENON}),\ }\href
  {\doibase 10.1103/PhysRevLett.122.141301} {\bibfield  {journal} {\bibinfo
  {journal} {Phys. Rev. Lett.}\ }\textbf {\bibinfo {volume} {122}},\ \bibinfo
  {pages} {141301} (\bibinfo {year} {2019}{\natexlab{a}})},\ \Eprint
  {http://arxiv.org/abs/1902.03234} {arXiv:1902.03234} \BibitemShut {NoStop}%
\bibitem [{\citenamefont {Baxter}\ \emph {et~al.}(2021)\citenamefont {Baxter}
  \emph {et~al.}}]{Baxter:2021pqo}%
  \BibitemOpen
  \bibfield  {author} {\bibinfo {author} {\bibfnamefont {D.}~\bibnamefont
  {Baxter}} \emph {et~al.},\ }\href {\doibase 10.1140/epjc/s10052-021-09655-y}
  {\bibfield  {journal} {\bibinfo  {journal} {Eur. Phys. J. C}\ }\textbf
  {\bibinfo {volume} {81}},\ \bibinfo {pages} {907} (\bibinfo {year} {2021})},\
  \Eprint {http://arxiv.org/abs/2105.00599} {arXiv:2105.00599 [hep-ex]}
  \BibitemShut {NoStop}%
\bibitem [{\citenamefont {Read}(2014)}]{Read:2014qva}%
  \BibitemOpen
  \bibfield  {author} {\bibinfo {author} {\bibfnamefont {J.}~\bibnamefont
  {Read}},\ }\href {\doibase 10.1088/0954-3899/41/6/063101} {\bibfield
  {journal} {\bibinfo  {journal} {J. Phys. G}\ }\textbf {\bibinfo {volume}
  {41}},\ \bibinfo {pages} {063101} (\bibinfo {year} {2014})},\ \Eprint
  {http://arxiv.org/abs/1404.1938} {arXiv:1404.1938} \BibitemShut {NoStop}%
\bibitem [{\citenamefont {Bramante}\ \emph {et~al.}(2019)\citenamefont
  {Bramante}, \citenamefont {Broerman}, \citenamefont {Kumar}, \citenamefont
  {Lang}, \citenamefont {Pospelov},\ and\ \citenamefont
  {Raj}}]{Bramante:2018tos}%
  \BibitemOpen
  \bibfield  {author} {\bibinfo {author} {\bibfnamefont {J.}~\bibnamefont
  {Bramante}}, \bibinfo {author} {\bibfnamefont {B.}~\bibnamefont {Broerman}},
  \bibinfo {author} {\bibfnamefont {J.}~\bibnamefont {Kumar}}, \bibinfo
  {author} {\bibfnamefont {R.~F.}\ \bibnamefont {Lang}}, \bibinfo {author}
  {\bibfnamefont {M.}~\bibnamefont {Pospelov}}, \ and\ \bibinfo {author}
  {\bibfnamefont {N.}~\bibnamefont {Raj}},\ }\href {\doibase
  10.1103/PhysRevD.99.083010} {\bibfield  {journal} {\bibinfo  {journal} {Phys.
  Rev. D}\ }\textbf {\bibinfo {volume} {99}},\ \bibinfo {pages} {083010}
  (\bibinfo {year} {2019})},\ \Eprint {http://arxiv.org/abs/1812.09325}
  {arXiv:1812.09325} \BibitemShut {NoStop}%
\bibitem [{\citenamefont {Clark}\ \emph {et~al.}(2020)\citenamefont {Clark},
  \citenamefont {Depoian}, \citenamefont {Elshimy}, \citenamefont {Kopec},
  \citenamefont {Lang}, \citenamefont {Li},\ and\ \citenamefont
  {Qin}}]{Clark:2020mna}%
  \BibitemOpen
  \bibfield  {author} {\bibinfo {author} {\bibfnamefont {M.}~\bibnamefont
  {Clark}}, \bibinfo {author} {\bibfnamefont {A.}~\bibnamefont {Depoian}},
  \bibinfo {author} {\bibfnamefont {B.}~\bibnamefont {Elshimy}}, \bibinfo
  {author} {\bibfnamefont {A.}~\bibnamefont {Kopec}}, \bibinfo {author}
  {\bibfnamefont {R.~F.}\ \bibnamefont {Lang}}, \bibinfo {author}
  {\bibfnamefont {S.}~\bibnamefont {Li}}, \ and\ \bibinfo {author}
  {\bibfnamefont {J.}~\bibnamefont {Qin}},\ }\href {\doibase
  10.1103/PhysRevD.102.123026} {\bibfield  {journal} {\bibinfo  {journal}
  {Phys. Rev. D}\ }\textbf {\bibinfo {volume} {102}},\ \bibinfo {pages}
  {123026} (\bibinfo {year} {2020})},\ \Eprint
  {http://arxiv.org/abs/2009.07909} {arXiv:2009.07909} \BibitemShut {NoStop}%
\bibitem [{\citenamefont {Bramante}\ \emph {et~al.}(2018)\citenamefont
  {Bramante}, \citenamefont {Broerman}, \citenamefont {Lang},\ and\
  \citenamefont {Raj}}]{Bramante:2018qbc}%
  \BibitemOpen
  \bibfield  {author} {\bibinfo {author} {\bibfnamefont {J.}~\bibnamefont
  {Bramante}}, \bibinfo {author} {\bibfnamefont {B.}~\bibnamefont {Broerman}},
  \bibinfo {author} {\bibfnamefont {R.~F.}\ \bibnamefont {Lang}}, \ and\
  \bibinfo {author} {\bibfnamefont {N.}~\bibnamefont {Raj}},\ }\href {\doibase
  10.1103/PhysRevD.98.083516} {\bibfield  {journal} {\bibinfo  {journal} {Phys.
  Rev. D}\ }\textbf {\bibinfo {volume} {98}},\ \bibinfo {pages} {083516}
  (\bibinfo {year} {2018})},\ \Eprint {http://arxiv.org/abs/1803.08044}
  {arXiv:1803.08044} \BibitemShut {NoStop}%
\bibitem [{\citenamefont {Aprile}\ \emph {et~al.}(2018)\citenamefont {Aprile}
  \emph {et~al.}}]{XENON:2018voc}%
  \BibitemOpen
  \bibfield  {author} {\bibinfo {author} {\bibfnamefont {E.}~\bibnamefont
  {Aprile}} \emph {et~al.} (\bibinfo {collaboration} {XENON}),\ }\href
  {\doibase 10.1103/PhysRevLett.121.111302} {\bibfield  {journal} {\bibinfo
  {journal} {Phys. Rev. Lett.}\ }\textbf {\bibinfo {volume} {121}},\ \bibinfo
  {pages} {111302} (\bibinfo {year} {2018})},\ \Eprint
  {http://arxiv.org/abs/1805.12562} {arXiv:1805.12562} \BibitemShut {NoStop}%
\bibitem [{\citenamefont {Aprile}\ \emph {et~al.}(2017)\citenamefont {Aprile}
  \emph {et~al.}}]{XENON:2017lvq}%
  \BibitemOpen
  \bibfield  {author} {\bibinfo {author} {\bibfnamefont {E.}~\bibnamefont
  {Aprile}} \emph {et~al.} (\bibinfo {collaboration} {XENON}),\ }\href
  {\doibase 10.1140/epjc/s10052-017-5326-3} {\bibfield  {journal} {\bibinfo
  {journal} {Eur. Phys. J. C}\ }\textbf {\bibinfo {volume} {77}},\ \bibinfo
  {pages} {881} (\bibinfo {year} {2017})},\ \Eprint
  {http://arxiv.org/abs/1708.07051} {arXiv:1708.07051} \BibitemShut {NoStop}%
\bibitem [{\citenamefont {Tovey}\ \emph {et~al.}(2000)\citenamefont {Tovey},
  \citenamefont {Gaitskell}, \citenamefont {Gondolo}, \citenamefont
  {Ramachers},\ and\ \citenamefont {Roszkowski}}]{Tovey:2000mm}%
  \BibitemOpen
  \bibfield  {author} {\bibinfo {author} {\bibfnamefont {D.~R.}\ \bibnamefont
  {Tovey}}, \bibinfo {author} {\bibfnamefont {R.~J.}\ \bibnamefont
  {Gaitskell}}, \bibinfo {author} {\bibfnamefont {P.}~\bibnamefont {Gondolo}},
  \bibinfo {author} {\bibfnamefont {Y.~A.}\ \bibnamefont {Ramachers}}, \ and\
  \bibinfo {author} {\bibfnamefont {L.}~\bibnamefont {Roszkowski}},\ }\href
  {\doibase 10.1016/S0370-2693(00)00846-7} {\bibfield  {journal} {\bibinfo
  {journal} {Phys. Lett. B}\ }\textbf {\bibinfo {volume} {488}},\ \bibinfo
  {pages} {17} (\bibinfo {year} {2000})},\ \Eprint
  {http://arxiv.org/abs/hep-ph/0005041} {arXiv:hep-ph/0005041} \BibitemShut
  {NoStop}%
\bibitem [{\citenamefont {Freytsis}\ and\ \citenamefont
  {Ligeti}(2011)}]{Freytsis:2010ne}%
  \BibitemOpen
  \bibfield  {author} {\bibinfo {author} {\bibfnamefont {M.}~\bibnamefont
  {Freytsis}}\ and\ \bibinfo {author} {\bibfnamefont {Z.}~\bibnamefont
  {Ligeti}},\ }\href {\doibase 10.1103/PhysRevD.83.115009} {\bibfield
  {journal} {\bibinfo  {journal} {Phys. Rev. D}\ }\textbf {\bibinfo {volume}
  {83}},\ \bibinfo {pages} {115009} (\bibinfo {year} {2011})},\ \Eprint
  {http://arxiv.org/abs/1012.5317} {arXiv:1012.5317} \BibitemShut {NoStop}%
\bibitem [{\citenamefont {Helm}(1956)}]{Helm:1956zz}%
  \BibitemOpen
  \bibfield  {author} {\bibinfo {author} {\bibfnamefont {R.~H.}\ \bibnamefont
  {Helm}},\ }\href {\doibase 10.1103/PhysRev.104.1466} {\bibfield  {journal}
  {\bibinfo  {journal} {Phys. Rev.}\ }\textbf {\bibinfo {volume} {104}},\
  \bibinfo {pages} {1466} (\bibinfo {year} {1956})}\BibitemShut {NoStop}%
\bibitem [{\citenamefont {Hardy}\ \emph {et~al.}(2015)\citenamefont {Hardy},
  \citenamefont {Lasenby}, \citenamefont {March-Russell},\ and\ \citenamefont
  {West}}]{Hardy:2015boa}%
  \BibitemOpen
  \bibfield  {author} {\bibinfo {author} {\bibfnamefont {E.}~\bibnamefont
  {Hardy}}, \bibinfo {author} {\bibfnamefont {R.}~\bibnamefont {Lasenby}},
  \bibinfo {author} {\bibfnamefont {J.}~\bibnamefont {March-Russell}}, \ and\
  \bibinfo {author} {\bibfnamefont {S.~M.}\ \bibnamefont {West}},\ }\href
  {\doibase 10.1007/JHEP07(2015)133} {\bibfield  {journal} {\bibinfo  {journal}
  {JHEP}\ }\textbf {\bibinfo {volume} {07}},\ \bibinfo {pages} {133} (\bibinfo
  {year} {2015})},\ \Eprint {http://arxiv.org/abs/1504.05419}
  {arXiv:1504.05419} \BibitemShut {NoStop}%
\bibitem [{\citenamefont {Vietze}\ \emph {et~al.}(2015)\citenamefont {Vietze},
  \citenamefont {Klos}, \citenamefont {Men\'endez}, \citenamefont {Haxton},\
  and\ \citenamefont {Schwenk}}]{Vietze:2014vsa}%
  \BibitemOpen
  \bibfield  {author} {\bibinfo {author} {\bibfnamefont {L.}~\bibnamefont
  {Vietze}}, \bibinfo {author} {\bibfnamefont {P.}~\bibnamefont {Klos}},
  \bibinfo {author} {\bibfnamefont {J.}~\bibnamefont {Men\'endez}}, \bibinfo
  {author} {\bibfnamefont {W.~C.}\ \bibnamefont {Haxton}}, \ and\ \bibinfo
  {author} {\bibfnamefont {A.}~\bibnamefont {Schwenk}},\ }\href {\doibase
  10.1103/PhysRevD.91.043520} {\bibfield  {journal} {\bibinfo  {journal} {Phys.
  Rev. D}\ }\textbf {\bibinfo {volume} {91}},\ \bibinfo {pages} {043520}
  (\bibinfo {year} {2015})},\ \Eprint {http://arxiv.org/abs/1412.6091}
  {arXiv:1412.6091} \BibitemShut {NoStop}%
\bibitem [{\citenamefont {Fitzpatrick}\ \emph {et~al.}(2013)\citenamefont
  {Fitzpatrick}, \citenamefont {Haxton}, \citenamefont {Katz}, \citenamefont
  {Lubbers},\ and\ \citenamefont {Xu}}]{Fitzpatrick:2012ix}%
  \BibitemOpen
  \bibfield  {author} {\bibinfo {author} {\bibfnamefont {A.~L.}\ \bibnamefont
  {Fitzpatrick}}, \bibinfo {author} {\bibfnamefont {W.}~\bibnamefont {Haxton}},
  \bibinfo {author} {\bibfnamefont {E.}~\bibnamefont {Katz}}, \bibinfo {author}
  {\bibfnamefont {N.}~\bibnamefont {Lubbers}}, \ and\ \bibinfo {author}
  {\bibfnamefont {Y.}~\bibnamefont {Xu}},\ }\href {\doibase
  10.1088/1475-7516/2013/02/004} {\bibfield  {journal} {\bibinfo  {journal}
  {JCAP}\ }\textbf {\bibinfo {volume} {02}},\ \bibinfo {pages} {004} (\bibinfo
  {year} {2013})},\ \Eprint {http://arxiv.org/abs/1203.3542} {arXiv:1203.3542}
  \BibitemShut {NoStop}%
\bibitem [{\citenamefont {Hoferichter}\ \emph {et~al.}(2019)\citenamefont
  {Hoferichter}, \citenamefont {Klos}, \citenamefont {Men\'endez},\ and\
  \citenamefont {Schwenk}}]{Hoferichter:2018acd}%
  \BibitemOpen
  \bibfield  {author} {\bibinfo {author} {\bibfnamefont {M.}~\bibnamefont
  {Hoferichter}}, \bibinfo {author} {\bibfnamefont {P.}~\bibnamefont {Klos}},
  \bibinfo {author} {\bibfnamefont {J.}~\bibnamefont {Men\'endez}}, \ and\
  \bibinfo {author} {\bibfnamefont {A.}~\bibnamefont {Schwenk}},\ }\href
  {\doibase 10.1103/PhysRevD.99.055031} {\bibfield  {journal} {\bibinfo
  {journal} {Phys. Rev. D}\ }\textbf {\bibinfo {volume} {99}},\ \bibinfo
  {pages} {055031} (\bibinfo {year} {2019})},\ \Eprint
  {http://arxiv.org/abs/1812.05617} {arXiv:1812.05617} \BibitemShut {NoStop}%
\bibitem [{\citenamefont {Engel}\ \emph {et~al.}(1992)\citenamefont {Engel},
  \citenamefont {Pittel},\ and\ \citenamefont {Vogel}}]{Engel:1992bf}%
  \BibitemOpen
  \bibfield  {author} {\bibinfo {author} {\bibfnamefont {J.}~\bibnamefont
  {Engel}}, \bibinfo {author} {\bibfnamefont {S.}~\bibnamefont {Pittel}}, \
  and\ \bibinfo {author} {\bibfnamefont {P.}~\bibnamefont {Vogel}},\ }\href
  {\doibase 10.1142/S0218301392000023} {\bibfield  {journal} {\bibinfo
  {journal} {Int. J. Mod. Phys. E}\ }\textbf {\bibinfo {volume} {1}},\ \bibinfo
  {pages} {1} (\bibinfo {year} {1992})}\BibitemShut {NoStop}%
\bibitem [{\citenamefont {Menendez}\ \emph {et~al.}(2012)\citenamefont
  {Menendez}, \citenamefont {Gazit},\ and\ \citenamefont
  {Schwenk}}]{Menendez:2012tm}%
  \BibitemOpen
  \bibfield  {author} {\bibinfo {author} {\bibfnamefont {J.}~\bibnamefont
  {Menendez}}, \bibinfo {author} {\bibfnamefont {D.}~\bibnamefont {Gazit}}, \
  and\ \bibinfo {author} {\bibfnamefont {A.}~\bibnamefont {Schwenk}},\ }\href
  {\doibase 10.1103/PhysRevD.86.103511} {\bibfield  {journal} {\bibinfo
  {journal} {Phys. Rev. D}\ }\textbf {\bibinfo {volume} {86}},\ \bibinfo
  {pages} {103511} (\bibinfo {year} {2012})},\ \Eprint
  {http://arxiv.org/abs/1208.1094} {arXiv:1208.1094} \BibitemShut {NoStop}%
\bibitem [{\citenamefont {Klos}\ \emph {et~al.}(2013)\citenamefont {Klos},
  \citenamefont {Men\'endez}, \citenamefont {Gazit},\ and\ \citenamefont
  {Schwenk}}]{Klos:2013rwa}%
  \BibitemOpen
  \bibfield  {author} {\bibinfo {author} {\bibfnamefont {P.}~\bibnamefont
  {Klos}}, \bibinfo {author} {\bibfnamefont {J.}~\bibnamefont {Men\'endez}},
  \bibinfo {author} {\bibfnamefont {D.}~\bibnamefont {Gazit}}, \ and\ \bibinfo
  {author} {\bibfnamefont {A.}~\bibnamefont {Schwenk}},\ }\href {\doibase
  10.1103/PhysRevD.88.083516} {\bibfield  {journal} {\bibinfo  {journal} {Phys.
  Rev. D}\ }\textbf {\bibinfo {volume} {88}},\ \bibinfo {pages} {083516}
  (\bibinfo {year} {2013})},\ \bibinfo {note} {[Erratum: Phys.Rev.D 89, 029901
  (2014)]},\ \Eprint {http://arxiv.org/abs/1304.7684} {arXiv:1304.7684}
  \BibitemShut {NoStop}%
\bibitem [{\citenamefont {Ressell}\ and\ \citenamefont
  {Dean}(1997)}]{Ressell:1997kx}%
  \BibitemOpen
  \bibfield  {author} {\bibinfo {author} {\bibfnamefont {M.~T.}\ \bibnamefont
  {Ressell}}\ and\ \bibinfo {author} {\bibfnamefont {D.~J.}\ \bibnamefont
  {Dean}},\ }\href {\doibase 10.1103/PhysRevC.56.535} {\bibfield  {journal}
  {\bibinfo  {journal} {Phys. Rev. C}\ }\textbf {\bibinfo {volume} {56}},\
  \bibinfo {pages} {535} (\bibinfo {year} {1997})},\ \Eprint
  {http://arxiv.org/abs/hep-ph/9702290} {arXiv:hep-ph/9702290} \BibitemShut
  {NoStop}%
\bibitem [{\citenamefont {Toivanen}\ \emph {et~al.}(2009)\citenamefont
  {Toivanen}, \citenamefont {Kortelainen}, \citenamefont {Suhonen},\ and\
  \citenamefont {Toivanen}}]{Toivanen:2009zza}%
  \BibitemOpen
  \bibfield  {author} {\bibinfo {author} {\bibfnamefont {P.}~\bibnamefont
  {Toivanen}}, \bibinfo {author} {\bibfnamefont {M.}~\bibnamefont
  {Kortelainen}}, \bibinfo {author} {\bibfnamefont {J.}~\bibnamefont
  {Suhonen}}, \ and\ \bibinfo {author} {\bibfnamefont {J.}~\bibnamefont
  {Toivanen}},\ }\href {\doibase 10.1103/PhysRevC.79.044302} {\bibfield
  {journal} {\bibinfo  {journal} {Phys. Rev. C}\ }\textbf {\bibinfo {volume}
  {79}},\ \bibinfo {pages} {044302} (\bibinfo {year} {2009})}\BibitemShut
  {NoStop}%
\bibitem [{\citenamefont {Bednyakov}\ and\ \citenamefont
  {Simkovic}(2005)}]{Bednyakov:2004xq}%
  \BibitemOpen
  \bibfield  {author} {\bibinfo {author} {\bibfnamefont {V.~A.}\ \bibnamefont
  {Bednyakov}}\ and\ \bibinfo {author} {\bibfnamefont {F.}~\bibnamefont
  {Simkovic}},\ }\href@noop {} {\bibfield  {journal} {\bibinfo  {journal}
  {Phys. Part. Nucl.}\ }\textbf {\bibinfo {volume} {36}},\ \bibinfo {pages}
  {131} (\bibinfo {year} {2005})},\ \Eprint
  {http://arxiv.org/abs/hep-ph/0406218} {arXiv:hep-ph/0406218} \BibitemShut
  {NoStop}%
\bibitem [{\citenamefont {{XENON
  Collaboration}}(2018)}]{xenon_collaboration_2018_1195785}%
  \BibitemOpen
  \bibfield  {author} {\bibinfo {author} {\bibnamefont {{XENON
  Collaboration}}},\ }\href {\doibase 10.5281/zenodo.1195785} {\enquote
  {\bibinfo {title} {The pax data processor v6.8.0},}\ } (\bibinfo {year}
  {2018})\BibitemShut {NoStop}%
\bibitem [{\citenamefont {Aalbers}\ \emph {et~al.}(2022)\citenamefont
  {Aalbers}, \citenamefont {Pelssers}, \citenamefont {Angevaare},\ and\
  \citenamefont {Morå}}]{jelle_aalbers_2022_7041453}%
  \BibitemOpen
  \bibfield  {author} {\bibinfo {author} {\bibfnamefont {J.}~\bibnamefont
  {Aalbers}}, \bibinfo {author} {\bibfnamefont {B.}~\bibnamefont {Pelssers}},
  \bibinfo {author} {\bibfnamefont {J.~R.}\ \bibnamefont {Angevaare}}, \ and\
  \bibinfo {author} {\bibfnamefont {K.~D.}\ \bibnamefont {Morå}},\ }\href
  {\doibase 10.5281/zenodo.7041453} {\enquote {\bibinfo {title} {Wimprates},}\
  } (\bibinfo {year} {2022})\BibitemShut {NoStop}%
\bibitem [{\citenamefont {Szydagis}\ \emph {et~al.}(2022)\citenamefont
  {Szydagis}, \citenamefont {Balajthy}, \citenamefont {Block}, \citenamefont
  {Brodsky}, \citenamefont {Cutter}, \citenamefont {Farrell}, \citenamefont
  {Huang}, \citenamefont {Kozlova}, \citenamefont {Lenardo}, \citenamefont
  {Manalaysay}, \citenamefont {McKinsey}, \citenamefont {Mooney}, \citenamefont
  {Mueller}, \citenamefont {Ni}, \citenamefont {Rischbieter}, \citenamefont
  {Tripathi}, \citenamefont {Tunnell}, \citenamefont {Velan},\ and\
  \citenamefont {Zhao}}]{szydagis_m_2022_6808388}%
  \BibitemOpen
  \bibfield  {author} {\bibinfo {author} {\bibfnamefont {M.}~\bibnamefont
  {Szydagis}}, \bibinfo {author} {\bibfnamefont {J.}~\bibnamefont {Balajthy}},
  \bibinfo {author} {\bibfnamefont {G.}~\bibnamefont {Block}}, \bibinfo
  {author} {\bibfnamefont {J.}~\bibnamefont {Brodsky}}, \bibinfo {author}
  {\bibfnamefont {J.}~\bibnamefont {Cutter}}, \bibinfo {author} {\bibfnamefont
  {S.}~\bibnamefont {Farrell}}, \bibinfo {author} {\bibfnamefont
  {J.}~\bibnamefont {Huang}}, \bibinfo {author} {\bibfnamefont
  {E.}~\bibnamefont {Kozlova}}, \bibinfo {author} {\bibfnamefont
  {B.}~\bibnamefont {Lenardo}}, \bibinfo {author} {\bibfnamefont
  {A.}~\bibnamefont {Manalaysay}}, \bibinfo {author} {\bibfnamefont
  {D.}~\bibnamefont {McKinsey}}, \bibinfo {author} {\bibfnamefont
  {M.}~\bibnamefont {Mooney}}, \bibinfo {author} {\bibfnamefont
  {J.}~\bibnamefont {Mueller}}, \bibinfo {author} {\bibfnamefont
  {K.}~\bibnamefont {Ni}}, \bibinfo {author} {\bibfnamefont {G.}~\bibnamefont
  {Rischbieter}}, \bibinfo {author} {\bibfnamefont {M.}~\bibnamefont
  {Tripathi}}, \bibinfo {author} {\bibfnamefont {C.}~\bibnamefont {Tunnell}},
  \bibinfo {author} {\bibfnamefont {V.}~\bibnamefont {Velan}}, \ and\ \bibinfo
  {author} {\bibfnamefont {Z.}~\bibnamefont {Zhao}},\ }\href {\doibase
  10.5281/zenodo.6808388} {\enquote {\bibinfo {title} {Noble element simulation
  technique},}\ } (\bibinfo {year} {2022})\BibitemShut {NoStop}%
\bibitem [{\citenamefont {Aprile}\ \emph
  {et~al.}(2019{\natexlab{b}})\citenamefont {Aprile} \emph
  {et~al.}}]{XENON:2019izt}%
  \BibitemOpen
  \bibfield  {author} {\bibinfo {author} {\bibfnamefont {E.}~\bibnamefont
  {Aprile}} \emph {et~al.} (\bibinfo {collaboration} {XENON}),\ }\href
  {\doibase 10.1103/PhysRevD.99.112009} {\bibfield  {journal} {\bibinfo
  {journal} {Phys. Rev. D}\ }\textbf {\bibinfo {volume} {99}},\ \bibinfo
  {pages} {112009} (\bibinfo {year} {2019}{\natexlab{b}})},\ \Eprint
  {http://arxiv.org/abs/1902.11297} {arXiv:1902.11297} \BibitemShut {NoStop}%
\bibitem [{\citenamefont {Baudis}\ \emph {et~al.}(2013)\citenamefont {Baudis},
  \citenamefont {Kessler}, \citenamefont {Klos}, \citenamefont {Lang},
  \citenamefont {Men\'endez}, \citenamefont {Reichard},\ and\ \citenamefont
  {Schwenk}}]{Baudis:2013bba}%
  \BibitemOpen
  \bibfield  {author} {\bibinfo {author} {\bibfnamefont {L.}~\bibnamefont
  {Baudis}}, \bibinfo {author} {\bibfnamefont {G.}~\bibnamefont {Kessler}},
  \bibinfo {author} {\bibfnamefont {P.}~\bibnamefont {Klos}}, \bibinfo {author}
  {\bibfnamefont {R.~F.}\ \bibnamefont {Lang}}, \bibinfo {author}
  {\bibfnamefont {J.}~\bibnamefont {Men\'endez}}, \bibinfo {author}
  {\bibfnamefont {S.}~\bibnamefont {Reichard}}, \ and\ \bibinfo {author}
  {\bibfnamefont {A.}~\bibnamefont {Schwenk}},\ }\href {\doibase
  10.1103/PhysRevD.88.115014} {\bibfield  {journal} {\bibinfo  {journal} {Phys.
  Rev. D}\ }\textbf {\bibinfo {volume} {88}},\ \bibinfo {pages} {115014}
  (\bibinfo {year} {2013})},\ \Eprint {http://arxiv.org/abs/1309.0825}
  {arXiv:1309.0825} \BibitemShut {NoStop}%
\bibitem [{\citenamefont {Aprile}\ \emph
  {et~al.}(2014{\natexlab{a}})\citenamefont {Aprile} \emph
  {et~al.}}]{XENON100:2013wdu}%
  \BibitemOpen
  \bibfield  {author} {\bibinfo {author} {\bibfnamefont {E.}~\bibnamefont
  {Aprile}} \emph {et~al.} (\bibinfo {collaboration} {XENON100}),\ }\href
  {\doibase 10.1088/0954-3899/41/3/035201} {\bibfield  {journal} {\bibinfo
  {journal} {J. Phys. G}\ }\textbf {\bibinfo {volume} {41}},\ \bibinfo {pages}
  {035201} (\bibinfo {year} {2014}{\natexlab{a}})},\ \Eprint
  {http://arxiv.org/abs/1311.1088} {arXiv:1311.1088} \BibitemShut {NoStop}%
\bibitem [{\citenamefont {Aprile}\ \emph
  {et~al.}(2014{\natexlab{b}})\citenamefont {Aprile} \emph
  {et~al.}}]{Aprile:2014zvw}%
  \BibitemOpen
  \bibfield  {author} {\bibinfo {author} {\bibfnamefont {E.}~\bibnamefont
  {Aprile}} \emph {et~al.} (\bibinfo {collaboration} {XENON1T}),\ }\href
  {\doibase 10.1088/1748-0221/9/11/P11006} {\bibfield  {journal} {\bibinfo
  {journal} {JINST}\ }\textbf {\bibinfo {volume} {9}},\ \bibinfo {pages}
  {P11006} (\bibinfo {year} {2014}{\natexlab{b}})},\ \Eprint
  {http://arxiv.org/abs/1406.2374} {arXiv:1406.2374} \BibitemShut {NoStop}%
\bibitem [{\citenamefont {Aglietta}\ \emph {et~al.}(1998)\citenamefont
  {Aglietta} \emph {et~al.}}]{LVD:1998lir}%
  \BibitemOpen
  \bibfield  {author} {\bibinfo {author} {\bibfnamefont {M.}~\bibnamefont
  {Aglietta}} \emph {et~al.} (\bibinfo {collaboration} {LVD}),\ }\href
  {\doibase 10.1103/PhysRevD.58.092005} {\bibfield  {journal} {\bibinfo
  {journal} {Phys. Rev. D}\ }\textbf {\bibinfo {volume} {58}},\ \bibinfo
  {pages} {092005} (\bibinfo {year} {1998})},\ \Eprint
  {http://arxiv.org/abs/hep-ex/9806001} {arXiv:hep-ex/9806001} \BibitemShut
  {NoStop}%
\bibitem [{\citenamefont {Aprile}\ \emph
  {et~al.}(2019{\natexlab{c}})\citenamefont {Aprile} \emph
  {et~al.}}]{XENON:2019ykp}%
  \BibitemOpen
  \bibfield  {author} {\bibinfo {author} {\bibfnamefont {E.}~\bibnamefont
  {Aprile}} \emph {et~al.} (\bibinfo {collaboration} {XENON}),\ }\href
  {\doibase 10.1103/PhysRevD.100.052014} {\bibfield  {journal} {\bibinfo
  {journal} {Phys. Rev. D}\ }\textbf {\bibinfo {volume} {100}},\ \bibinfo
  {pages} {052014} (\bibinfo {year} {2019}{\natexlab{c}})},\ \Eprint
  {http://arxiv.org/abs/1906.04717} {arXiv:1906.04717} \BibitemShut {NoStop}%
\bibitem [{\citenamefont {Jenks}(1967)}]{Jenks1967TheDM}%
  \BibitemOpen
  \bibfield  {author} {\bibinfo {author} {\bibfnamefont {G.~F.}\ \bibnamefont
  {Jenks}},\ }in\ \href@noop {} {\emph {\bibinfo {booktitle} {International
  Yearbook of Cartography}}},\ \bibinfo {series and number} {\bibinfo {number}
  {7}}\ (\bibinfo {year} {1967})\ pp.\ \bibinfo {pages}
  {{186--190}}\BibitemShut {NoStop}%
\bibitem [{\citenamefont {Alvis}\ \emph {et~al.}(2018)\citenamefont {Alvis}
  \emph {et~al.}}]{Majorana:2018gib}%
  \BibitemOpen
  \bibfield  {author} {\bibinfo {author} {\bibfnamefont {S.~I.}\ \bibnamefont
  {Alvis}} \emph {et~al.} (\bibinfo {collaboration} {Majorana}),\ }\href
  {\doibase 10.1103/PhysRevLett.120.211804} {\bibfield  {journal} {\bibinfo
  {journal} {Phys. Rev. Lett.}\ }\textbf {\bibinfo {volume} {120}},\ \bibinfo
  {pages} {211804} (\bibinfo {year} {2018})},\ \Eprint
  {http://arxiv.org/abs/1801.10145} {arXiv:1801.10145 [hep-ex]} \BibitemShut
  {NoStop}%
\bibitem [{\citenamefont {Adhikari}\ \emph {et~al.}(2022)\citenamefont
  {Adhikari} \emph {et~al.}}]{DEAPCollaboration:2021raj}%
  \BibitemOpen
  \bibfield  {author} {\bibinfo {author} {\bibfnamefont {P.}~\bibnamefont
  {Adhikari}} \emph {et~al.} (\bibinfo {collaboration} {DEAP Collaboration}),\
  }\href {\doibase 10.1103/PhysRevLett.128.011801} {\bibfield  {journal}
  {\bibinfo  {journal} {Phys. Rev. Lett.}\ }\textbf {\bibinfo {volume} {128}},\
  \bibinfo {pages} {011801} (\bibinfo {year} {2022})},\ \Eprint
  {http://arxiv.org/abs/2108.09405} {arXiv:2108.09405} \BibitemShut {NoStop}%
\bibitem [{\citenamefont {Feldman}\ and\ \citenamefont
  {Cousins}(1998)}]{Feldman:1997qc}%
  \BibitemOpen
  \bibfield  {author} {\bibinfo {author} {\bibfnamefont {G.~J.}\ \bibnamefont
  {Feldman}}\ and\ \bibinfo {author} {\bibfnamefont {R.~D.}\ \bibnamefont
  {Cousins}},\ }\href {\doibase 10.1103/PhysRevD.57.3873} {\bibfield  {journal}
  {\bibinfo  {journal} {Phys. Rev. D}\ }\textbf {\bibinfo {volume} {57}},\
  \bibinfo {pages} {3873} (\bibinfo {year} {1998})},\ \Eprint
  {http://arxiv.org/abs/physics/9711021} {arXiv:physics/9711021} \BibitemShut
  {NoStop}%
\bibitem [{\citenamefont {Broerman}(2022)}]{Broerman:2022jtj}%
  \BibitemOpen
  \bibfield  {author} {\bibinfo {author} {\bibfnamefont {B.}~\bibnamefont
  {Broerman}},\ }\emph {\bibinfo {title} {\textrm{New ideas for tonne-scale
  bubble chambers and a search for superheavy dark matter with PICO-60}}},\
  \href@noop {} {Ph.D. thesis},\ \bibinfo  {school} {Queen's U., Kingston}
  (\bibinfo {year} {2022})\BibitemShut {NoStop}%
\bibitem [{\citenamefont {Acevedo}\ \emph {et~al.}(2021)\citenamefont
  {Acevedo}, \citenamefont {Bramante},\ and\ \citenamefont
  {Goodman}}]{Acevedo:2021tbl}%
  \BibitemOpen
  \bibfield  {author} {\bibinfo {author} {\bibfnamefont {J.~F.}\ \bibnamefont
  {Acevedo}}, \bibinfo {author} {\bibfnamefont {J.}~\bibnamefont {Bramante}}, \
  and\ \bibinfo {author} {\bibfnamefont {A.}~\bibnamefont {Goodman}}\
  }(\bibinfo {year} {2021})\ \Eprint {http://arxiv.org/abs/2105.06473}
  {arXiv:2105.06473} \BibitemShut {NoStop}%
\bibitem [{\citenamefont {Bhoonah}\ \emph {et~al.}(2021)\citenamefont
  {Bhoonah}, \citenamefont {Bramante}, \citenamefont {Courtman},\ and\
  \citenamefont {Song}}]{Bhoonah:2020fys}%
  \BibitemOpen
  \bibfield  {author} {\bibinfo {author} {\bibfnamefont {A.}~\bibnamefont
  {Bhoonah}}, \bibinfo {author} {\bibfnamefont {J.}~\bibnamefont {Bramante}},
  \bibinfo {author} {\bibfnamefont {B.}~\bibnamefont {Courtman}}, \ and\
  \bibinfo {author} {\bibfnamefont {N.}~\bibnamefont {Song}},\ }\href {\doibase
  10.1103/PhysRevD.103.103001} {\bibfield  {journal} {\bibinfo  {journal}
  {Phys. Rev. D}\ }\textbf {\bibinfo {volume} {103}},\ \bibinfo {pages}
  {103001} (\bibinfo {year} {2021})},\ \Eprint
  {http://arxiv.org/abs/2012.13406} {arXiv:2012.13406} \BibitemShut {NoStop}%
\bibitem [{\citenamefont {Aprile}\ \emph {et~al.}(2020)\citenamefont {Aprile}
  \emph {et~al.}}]{XENON:2020kmp}%
  \BibitemOpen
  \bibfield  {author} {\bibinfo {author} {\bibfnamefont {E.}~\bibnamefont
  {Aprile}} \emph {et~al.} (\bibinfo {collaboration} {XENON}),\ }\href
  {\doibase 10.1088/1475-7516/2020/11/031} {\bibfield  {journal} {\bibinfo
  {journal} {JCAP}\ }\textbf {\bibinfo {volume} {11}},\ \bibinfo {pages} {031}
  (\bibinfo {year} {2020})},\ \Eprint {http://arxiv.org/abs/2007.08796}
  {arXiv:2007.08796} \BibitemShut {NoStop}%
\bibitem [{\citenamefont {Akerib}\ \emph {et~al.}(2020)\citenamefont {Akerib}
  \emph {et~al.}}]{LZ:2018qzl}%
  \BibitemOpen
  \bibfield  {author} {\bibinfo {author} {\bibfnamefont {D.~S.}\ \bibnamefont
  {Akerib}} \emph {et~al.} (\bibinfo {collaboration} {LZ}),\ }\href {\doibase
  10.1103/PhysRevD.101.052002} {\bibfield  {journal} {\bibinfo  {journal}
  {Phys. Rev. D}\ }\textbf {\bibinfo {volume} {101}},\ \bibinfo {pages}
  {052002} (\bibinfo {year} {2020})},\ \Eprint
  {http://arxiv.org/abs/1802.06039} {arXiv:1802.06039} \BibitemShut {NoStop}%
\bibitem [{\citenamefont {Aalseth}\ \emph {et~al.}(2018)\citenamefont {Aalseth}
  \emph {et~al.}}]{DarkSide-20k:2017zyg}%
  \BibitemOpen
  \bibfield  {author} {\bibinfo {author} {\bibfnamefont {C.~E.}\ \bibnamefont
  {Aalseth}} \emph {et~al.} (\bibinfo {collaboration} {DarkSide-20k}),\ }\href
  {\doibase 10.1140/epjp/i2018-11973-4} {\bibfield  {journal} {\bibinfo
  {journal} {Eur. Phys. J. Plus}\ }\textbf {\bibinfo {volume} {133}},\ \bibinfo
  {pages} {131} (\bibinfo {year} {2018})},\ \Eprint
  {http://arxiv.org/abs/1707.08145} {arXiv:1707.08145} \BibitemShut {NoStop}%
\bibitem [{\citenamefont {Aalbers}\ \emph {et~al.}(2016)\citenamefont {Aalbers}
  \emph {et~al.}}]{DARWIN:2016hyl}%
  \BibitemOpen
  \bibfield  {author} {\bibinfo {author} {\bibfnamefont {J.}~\bibnamefont
  {Aalbers}} \emph {et~al.} (\bibinfo {collaboration} {DARWIN}),\ }\href
  {\doibase 10.1088/1475-7516/2016/11/017} {\bibfield  {journal} {\bibinfo
  {journal} {JCAP}\ }\textbf {\bibinfo {volume} {11}},\ \bibinfo {pages} {017}
  (\bibinfo {year} {2016})},\ \Eprint {http://arxiv.org/abs/1606.07001}
  {arXiv:1606.07001 [astro-ph.IM]} \BibitemShut {NoStop}%
\bibitem [{\citenamefont {Aalbers}\ \emph {et~al.}(2023)\citenamefont {Aalbers}
  \emph {et~al.}}]{Aalbers:2022dzr}%
  \BibitemOpen
  \bibfield  {author} {\bibinfo {author} {\bibfnamefont {J.}~\bibnamefont
  {Aalbers}} \emph {et~al.},\ }\href {\doibase 10.1088/1361-6471/ac841a}
  {\bibfield  {journal} {\bibinfo  {journal} {J. Phys. G}\ }\textbf {\bibinfo
  {volume} {50}},\ \bibinfo {pages} {013001} (\bibinfo {year} {2023})},\
  \Eprint {http://arxiv.org/abs/2203.02309} {arXiv:2203.02309
  [physics.ins-det]} \BibitemShut {NoStop}%
\end{thebibliography}%

\end{document}